# Equation of State and Constitutive Models for Numerical Simulations of Dust Impacts on the Solar Probe

Report on Contract #949182
Johns Hopkins University Applied Physics Laboratory
Laurel, Maryland

Gerald I. Kerley, Consultant

August 2009

# Equation of State and Constitutive Models for Numerical Simulations of Dust Impacts on the Solar Probe

Report on Contract #949182
Johns Hopkins University Applied Physics Laboratory
Laurel, Maryland

Gerald I. Kerley, Consultant

August 2009

## ABSTRACT

This report presents new EOS and strength models for use in numerical hydrocode simulations of dust impacts on the NASA solar probe space vehicle. This spacecraft will be subjected to impact at velocities up to 300 km/s, producing pressures as high as 100 TPa and temperatures as high as 200 eV. Hence the material models must treat a variety of physical and chemical phenomena, including solid-solid transitions, melting and vaporization, chemical reactions, electronic excitation and ionization. The EOSPro code is used to develop tabular EOS that include these effects. The report discusses the theoretical methods used to create the new EOS tables and constitutive models for six materials—$Al_2O_3$, two porous carbon materials, fused $SiO_2$, a silicone elastomer, and germanium—which will be used in the thermal protection shield (TPS) and solar cells, the components most vulnerable to dust impacts. It also presents the results of hydrocode simulations of dust impacts on the TPS and on glass targets. It discusses the importance of radiation contributions to the EOS and the importance of reactions between $Al_2O_3$ and carbon during impacts. It makes recommendations for additional work, including experiments for testing the accuracy and reliability of the hydrocode simulations as well as for improving and calibrating the material models.

# EXECUTIVE SUMMARY

The Solar Probe Plus (SPP) spacecraft will have to survive and function properly when subjected to impacts by dust particles, especially when flying close to the sun. These particles, though small as very fine grains of sand, can do much damage because of their extremely high velocities. A 100 μm particle, moving at 100 km/s, will create a crater of diameter 0.2 cm and will damage surrounding material over a much larger region.

In order to ensure success of the mission, it is necessary to assess this dust impact damage and find ways to mitigate it. Experimental studies can and will be used, but terrestrial laboratories will not be able to replicate the impact conditions to which the spacecraft will be exposed. Therefore, numerical hydrocode simulations will play a central role in analysis of the dust impact problem.

The work described in this report was commissioned to support numerical simulations using CTH, a hydrocode developed and maintained at Sandia National Laboratories, Albuquerque, NM. The CTH code is well-known throughout the hydrodynamics and shock-wave community, has proven useful for a wide variety of applied problems, and has many tools for modeling the equation of state (EOS) and constitutive properties needed for the calculations. CTH had already been used in a dust impact study made prior to this project.

This report presents the results of some CTH simulations but is devoted primarily to the evaluation and development of the EOS and constitutive models. In order to make this report informative and useful to a wide audience, the discussions are focused primarily on concepts, not on equations and details that would be of interest only to experts in this field. Indeed, the report even attempts to offer insights into certain practical problems and issues that are often avoided in work published in "scholarly" journals. For example, Sec. 2 gives an overview of the entire computational approach, including both hydrocode and material modeling issues, some of which are seldom discussed in the literature.

A hypervelocity dust impact is essentially an explosion. A large amount of energy is deposited in a small region, in a small time, resulting in ejection of material to form a crater and propagation of a diverging shock wave into surrounding material. The pressures and temperatures produced by such an explosion cover a phenomenal range, from 100 TPa and $2 \times 10^6$ K at the point of impact all the way down to ambient conditions as the wave propagates outward.

Hence calculations of dust impacts and similar events require the use of "global" EOS and constitutive models that are valid over the entire range. The physical and chemical phenomena that govern the material response change dramatically with pressure and temperature. At the lowest stresses and temperatures, materials are solids, and their properties are affected by shear and tensile forces and by solid-solid phase transitions. At higher stresses and temperatures, materials melt; they no longer support shear deformations, and the description of their EOS behavior requires different models than those used for solids. Higher stresses and temperatures lead to dissociation and ionization, conditions which introduce new material properties and require new modeling techniques.

This report presents either new or improved material models for six materials—aluminum oxide, a carbon-carbon composite, a very low-density carbon foam, fused silica, a silicone elastomer adhesive, and germanium. These materials will be used in the thermal protection shield (TPS) and the solar cells, the two spacecraft components most vulnerable to dust impacts. The report also discusses hydrocode simulations of dust impacts on the TPS and on glass targets, the importance of radiation terms in the EOS, and the possibility of reactions between carbon and $Al_2O_3$ during dust impacts.

The work discussed in this report was carried out during the "Pre-Phase A" planning period of the SPP project. It represents a significant step forward; much was accomplished in a very short time. However, much remains to be done. The report discusses recommendations for follow-on work, including ideas for experimental studies.

# CONTENTS

# FIGURES

# 1. INTRODUCTION

## 1.1 Background

Solar Probe Plus (SPP) is a NASA program to fly a science spacecraft into the solar corona, to measure properties of the solar plasma and radiation [1]. During the final part of the mission, the vehicle will be subjected to numerous impacts of dust particles. It is expected that these particles will be as large as 100 μm and that impact velocities will be as high as 300 km/s [2][3].

The present study was commissioned in support of investigations, currently underway at JHU/APL, to assess the effects of dust impact on the SPP spacecraft. It is obviously important to protect the vehicle and its components from damage that could destroy their structural integrity or function. It is also necessary to ensure that mass ejected into the plasma surrounding the spacecraft does not appreciably affect the environment that the science experiments are supposed to measure.

Two parts of the vehicle are of particular interest—the thermal protection shield (TPS) and the solar cells [4].

- The TPS is a laminate of several materials that faces directly towards the sun. The first layer is a ceramic (currently $Al_2O_3$), which serves as a radiation reflector and thermal barrier. Other layers include a carbon-carbon composite and a low-density carbon foam. The temperature will vary from 1650K at the top of the TPS to 800K at the bottom. Impact velocities are expected to be as high as 50 km/s.

- The solar cells, which provide power for the spacecraft and experiments, is actively cooled so that the temperature will stay below 473K. They consist of a glass cover, an adhesive layer, germanium, more adhesive, kapton, and a metal substrate. The impact velocities are expected to be as high as 300 km/s.

## 1.2 Technical Approach

The CTH hydrocode [5]-[10] is being used to make numerical simulations of dust impacts on the TPS and solar cells. There are several reasons why CTH is well suited for this task. First, it has the capability to treat the high states of deformation that occur in these impacts. Second, it allows the use of tabular equations of state (EOS) that can model the physical and chemical phenomena that become important at such high impact velocities. Third, it has a number of options for treating material strength, fracture, and porosity, which also have significant effects on some of the predictions. Fourth, it can be used in both two and three spatial dimensions, allowing one to study both normal and oblique impacts.

The EOS models required for these simulations must be valid over an extremely wide range of densities and temperatures. The ultra-high impact velocities result in temperatures above $10^6$ K. All materials are completely dissociated and at least partially ionized at such temperatures, so those phenomena need to be included in the model. However, the pressure and temperature drop off rapidly as the shock produced by the impact diverges from the impact point. Therefore, the EOS must also predict realistic behavior at lower pressure-temperature levels, where phase transitions, melting, vaporization, and chemical reactions control the material response. At the lowest stress levels, on the outskirts of the diverging waves, the strength and fracture behavior can also be important.

The EOSPro code [11], which has the capability to treat all of the above phenomena, was used to generate the EOS models needed for these simulations. The EOSPro models are put into tabular form for use in CTH [9][10]. The CTH calculations also employ the p-alpha model for porosity, the von Mises and Steinberg models for strength, and the minimum stress model for fracture.

The purpose of the present study was to develop new or improved EOS and constitutive models for use in these CTH simulations and to assist, where needed, in setting up and running the simulations themselves. The following tasks were identified before and during the study:

- Evaluate existing EOS models for carbon, $Al_2O_3$ (alumina), and glass; improve where needed.
- Develop models for treating porous carbon materials (composite and foam).
- Examine effects of microstructure on the behavior of these carbon materials in dust impacts on TPS.
- Develop new EOS tables for germanium and the solar cell adhesive.
- Develop strength and fracture parameters for all materials, especially glass and $Al_2O_3$, for which surface fracture could be important.
- Modify all EOS to include radiation terms.
- Investigate effects of reactions between $Al_2O_3$ and carbon in dust impacts on TPS.

## 1.3  Overview of Report

The present study was carried out during the so-called "Pre-Phase A" planning period, my level of effort being 0.5 FTE for six months. The magnitude of the tasks outlined above vastly exceeded the time and level of effort available to work on them. In order that my work be as useful as possible to the project planners, my goals had to be limited to preliminary and exploratory efforts, the overall objective being to identify which problems were of greatest importance and what kind of work would be required in subsequent studies.

There is, of course, a danger in taking this approach, i.e., that the project planners will regard these problems to be solved and not understand the need to go beyond the preliminary studies. In writing this report, I have attempted to strike a balance between touting the very real progress that has been made on these tasks without ignoring the many things that remain to be done. I can only hope that this report will receive the careful reading it deserves.

The computational modeling of hypervelocity dust impacts requires the use of computer codes and methods that may not be familiar to some readers of this report. Therefore, Section 2 gives an overview of the approach that is used in this work, the hydrocode, the EOS models, and the constitutive models.

Section 3 discusses the contribution of radiation terms to the EOS. It is shown that these terms will not be important for the impact conditions presently expected.

Sections 4-7 discuss material models and dust impacts on the TPS.
- Section 4 discusses the EOS of aluminum oxide, a ceramic that is planned for use as a thermal barrier in the top layer of the TPS.
- Section 5 discusses EOS for two porous carbon materials that will comprise most of the TPS.
- Section 6 discusses hydrocode calculations for typical dust impacts on the TPS.
- Section 7 discusses the possibility of reactions between the aluminum oxide and carbon layers during a dust impact.

Sections 8-10 discuss material models and dust impacts on solar cells.
- Section 8 discusses the EOS and constitutive model for fused silica and glass, the top layer of the SPP solar cells.
- Section 9 discusses a new EOS for Sylgard, which is being used in the simulations as a substitute for the solar cell adhesive, DC93-500.
- Section 10 discusses a new EOS for germanium, also used in the solar cells.

Section 11 discusses the need for experimental data to test the material models and numerical predictions and possible options for obtaining such data.

Section 12 summarizes the results of the study and offers recommendations for future work.

## 2. COMPUTATIONAL APPROACH

Hypervelocity dust impacts generate very high temperatures and pressures, accompanied by very high strain rates and deformations. In order to understand and model their effects, it is necessary to employ tools that may not be familiar to some readers of this report. This section offers an overview of the approach that is used in such studies and how it differs from methods used in other problems.

### 2.1 Hydrocodes[1]

Analyses of structural loading at low strain rates often use finite element codes that employ the so-called "Lagrangian" method, in which the computational mesh is embedded in the material and deforms along with it. This approach can also be used at high strain rates and is generally regarded as the most accurate numerical method for small states of deformation. However, it is unable to handle high deformations that result in hopelessly tangled meshes. Attempts to correct for mesh tangling, e.g., by rezoning and/or discard of material, are inadequate for the hypervelocity impact problems considered in this report.

The most widely-used computational method for high-deformation problems is called "Eulerian," because it involves a computational mesh that is fixed in space. This method actually involves two computational steps—a Lagrangian step, in which the mesh deforms with the material, followed by a "rezone" step, in which the variables are mapped back onto the original spatial mesh. Continual rezoning solves the problem of tangled meshes, but it introduces numerical errors. It also creates "mixed cells," cells containing more than one material. Eulerian hydrocodes offer many options for treating these mixed cells, but none of them can be rigorously derived from first principles or demonstrated to be exact.

Other numerical methods, including ALE (arbitrary Lagrangian-Eulerian) and SPH (smoothed-particle hydrodynamics) are also available. However, these methods have not been used and tested as extensively as the the standard Eulerian approach has. Moreover, codes employing those methods generally lack certain features for modeling the material properties that are essential for studying hypervelocity impacts, as discussed below.

The present work employs CTH, an Eulerian hydrocode developed and maintained at Sandia National Laboratories, Albuquerque, NM [5]-[8]. CTH offers options for hydrocode simulations in one, two, and three spatial dimensions. It is widely used throughout the U.S. Departments of Energy and Defense. It has been chosen for this work partly because it is well-known and readily available to the participants in this project, but also because it offers more powerful options for modeling the EOS and constitutive properties than other hydrocodes [9][10].

---

1. In this report, the term "hydrocode" is short for "continuum mechanics code," i.e., it applies to a computer program that solve the mass, momentum, and energy conservation equations in solids as well as in fluids.

It is sometimes suggested that the best approach for simulating problems like those considered here would be to combine both Eulerian and Lagrangian codes. The Eulerian code would be used in the early stages of the problem, where there are large deformations. After the stresses drop, as shock waves diverge from the point of impact, the calculation would be "picked up" by a Lagrangian code and used to treat the details of structural response more carefully.

This idea looks good on paper but is remarkably difficult to implement in practice. It requires that two codes, which use different physical descriptions of reality, find a way to communicate with one another. This approach has been tried again and again, usually with only limited success. And this capability apparently does not exist for the dust impact problems addressed in this study.

It also turns out that the two codes typically use quite different material models. Most structural response codes lack the sophisticated EOS models available in CTH but [are thought to] offer better models for constitutive properties, at least in the lower stress regime. However, it should be noted that large deformations can *also* occur in the *low*-stress regime, where the Lagrangian codes do not perform well. This situation is particularly evident in the calculations of spall damage in glass, as discussed in Sec. 8.

## 2.2 Global EOS Models

The term "global EOS" was invented [12] to describe equations of state that give reasonable behavior over the widest possible range of densities and temperatures, including the extreme conditions encountered in nuclear explosions.[1]

It is important to understand that most of the EOS models available in hydrocodes are *not* global. For example, the well-known, popular, Mie-Grüneisen EOS, which can be constructed using a fit to shock Hugoniot data, is often thought to be suitable for almost any problem that involves shock waves. This belief is erroneous for a variety of reasons that are discussed elsewhere [9][10][13].

Truly global EOS models are required for calculations of hypervelocity dust impacts because the shock loading and unloading states cover an enormous range of densities and temperatures. For example, the impact of a fused silica particle on a silica target, at 300 km/s, will produce a pressure of $6.2 \times 10^4$ GPa and a temperature of $1.9 \times 10^6$ K at the point of impact. Matter under these conditions is com-pletely dissociated and highly ionized. As the shock wave diverges from the point of impact, the pressure and temperature produced in the target drop all the way down to the ambient state. Hence *the shock produces all pressure-temperature states consistent with a shock wave*, i.e., the entire Hugoniot. The physical and chemical phenomena that govern the material behavior are very different in the various pressure-temperature regimes. On the outskirts of the impact, the material behavior is determined by properties of the solid: compression, including phase transitions, shear phenomena, damage, and spall. Closer to the point of impact, the shocked material will be melted and strength will be unimportant. The effects of molecular dissociation, electronic excitation, and ionization all become progressively more important at points closer and closer to the point of impact.[1]

---

1. The term global EOS, as used here, does *not* include nuclear matter itself.

A dust impact also produces states well *off* the Hugoniot, as the shocked material expands. At high impact velocities, the material blown out of the crater is gas-like, i.e., does not consist of solid-like fragments, and it undergoes enormous expansion. However, release from the shock state can also generate important EOS phenomena even at relatively *low* pressures, particularly when phase transitions and chemical reactions are involved. Hence, *it is not sufficient for the EOS to match the Hugoniot*. It must also give a realistic description of the unloading behavior.

Given the above observations, it would appear desirable to model all of the important physical-chemical phenomena that govern material behavior when creating an EOS model. However, most of the EOS codes currently available do not have the necessary capabilities. Many hydrocode users are not aware of this fact, regarding all EOS codes as more or less the same, using the same principles and differing primarily in "bells and whistles."

This work uses the EOSPro code [11], which has capabilities that are not available—so far as I know—in any other code in the world. Features that are especially important to the present study are:

- The ability to model a material with a complicated phase diagram. In the present work, this feature is used in the EOS for several materials, germanium being the newest model.

- A realistic description of the liquid phase. The liquid model, which is not available in other EOS codes, has been used with good results in every kind of material that appears in the SPP dust impact study.

- A realistic description of molecular phenomena at the lower temperatures and the effects of dissociation at high temperatures. This feature is important in the models for Sylgard, fused silica, and aluminum oxide.

- A unique model for the electronic excitation and ionization of matter. This feature has been used in constructing some of the materials needed for the SPP study, especially carbon. Unfortunately, it has not been possible to employ it in the new EOS models because of time limitations.

---

1. Note that a shock wave cannot vaporize a material, because vaporization only occurs when a material expands. This process *can* occur on unloading from a shock state.

There is another point that is often overlooked or misunderstood: It is not possible to represent EOS surfaces having such complexity by any kind of analytic expression. EOSPro models are put into tabular form (Sesame tables [9]-[11]) for use in the CTH code. The CTH EOS package also has several features that are not available in other codes using tabular EOS—the ability to include porosity (see Sec. 5.2) and reactive behavior, and an interpolation scheme that "inverts" the EOS (finds temperature as a function of energy) exactly.

It is sometimes claimed that tabular EOS are computationally inefficient, that they require more computing time than analytic models. This belief, however widespread it may be, is completely false. Actual timing tests have demonstrated that the tabular methods are actually much *faster* than the analytic alternatives [13].

One final observation must be made. Hydrocode users often view the hydrocode and EOS models as a "package," assuming that a good code implies a good library of EOS models. The sad truth is that hydrocode developers typically devote little time and effort to the development of an EOS database. Having been responsible for the CTH EOS package for many years, I claim that there are still many deficiencies in the EOS library. The reason is that sufficient resources were never allocated to improve this crucial feature. The existing database is inadequate for the dust impact study and improved models have to be developed.

## 2.3  Constitutive Models

The term "constitutive model" is generally used to describe all aspects of the material model besides the EOS, especially the material strength (shear behavior) and spall (damage, fracture, etc.).

CTH offers an impressive number of sophisticated constitutive models [6]. However, I have not even attempted to use most of them in the present study, for three reasons. *First*, model parameters are not available for most of the materials needed for the SPP project. *Second*, there was not sufficient time to develop the model parameters, given all the other tasks to be considered. *Third*, and perhaps most important, it is by no means clear just how sophisticated the constitutive model needs to be to affect the dust impact predictions. (In previous work, the author has found that a relatively simple constitutive model sometimes gives satisfactory results when used with a good EOS [14]—the opposite of what is often claimed by constitutive modelers.)

In this study, the material strength behavior for most materials has been modeled using the basic "von Mises" option in CTH [6]; the material is described by three parameters, all taken to be constants—Poisson's ratio, the yield strength, and the melting temperature. Poisson's ratio is used to calculate the shear modulus, which is, in turn, used to describe the elastic deformation. The yield strength is used to define a yield surface, the onset of plastic flow, assuming the material to be isotropic. The melting temperature determines the point at which material strength vanishes.

The Steinberg-Guinan-Lund (SGL) option [15][16] offers an improvement to the simple model. In the present work, SGL has been used primarily because it allows one to represent the melting curve more accurately. This feature was used in the studies of impacts on fused silica, as discussed in Sec. 8. SGL also allows one to include pressure, temperature, and strain-rate dependence. T-dependence was used in the model for Ge (Sec. 10); the other options have not yet been exploited.

The constitutive model must also describe tensile failure, which results in fracture or spall. For the simulation of dust impacts, this issue is particularly important for predicting the behavior of the surface materials, where significant spall damage outside the central crater area can result in additional loss and/or damage of material, loss of thermal protection to lower layers, and increased vulnerability to subsequent impacts.

The simplest tensile failure model in CTH, the one employed in this study, uses a minimum stress or pressure criterion [6]. Each material is assigned a minimum tensile strength (a negative number). If the stress/pressure in a computational cell drops below this minimum, the code opens a void region in the cell, bringing the material back to zero stress/pressure. Davison, et al. [17][18], claim that this model, though extremely simplistic, gives reasonable predictions of spall damage in fused silica when subjected to high velocity impacts up to at least 10 km/s.

In the present study, the spall damage predictions for glass were found to depend on two parameters besides the material tensile strengths. The results were surprisingly different for stress-based and pressure-based fracture criteria. The pressure criterion is recommended in the CTH documentation, but the stress criterion was found to give much better results. The results also depend on an input parameter PFVOID, the tensile strength of a mixed cell containing void. The spall predictions were completely inaccurate unless PFVOID was set to a value higher than the material tensile strength, and the predictions depended to a disturbing degree on the actual value. See Sec. 8 for further discussion.

CTH also offers an option, the Johnson-Cook fracture model [19], that couples the treatment of shear and tensile strength. In brief, plastic flow creates damage that decreases the tensile strength of the material, while tensile failure degrades the shear strength. I have found this model to be effective in predicting penetration in iron [14]. Unfortunately, this model could not be used in the present study because of lack of experimental information and time to calibrate it.

## 2.4 Code "Validation"

CTH and other hydrocodes can generate impressive results, including visualizations that look almost like the real thing. However, it is reasonable, indeed necessary, to ask: How accurate are these simulations? How much can they be trusted? It is astoundingly difficult to answer these questions.

It is common to find, in the literature, theorists who claim that their models have been "validated." This term is intended to convey the impression that the models have been exhaustively tested and shown to be accurate, that they can be trusted. Such claims generally do not bear up under scrutiny and should not be taken seriously. Use of the term "validated" may even be a clue that the author is trying to bluff, to avoid the difficult task of evaluating the accuracy of his model.

In the software industry, the term "validation" has a more restricted meaning, i.e., verification that a code performs as claimed in the documentation (assuming this documentation actually exists). In its simplest form, validation means that the code does not contain logical errors, e.g., replacement of a + sign with a - sign, or, more likely something more involved and harder to find. But elimination of all logical errors does not necessarily mean that the code gives accurate results.

For use in practical applications, it is far more important that the code be accurate than use pristine logic. One will even be willing to accept some bugs, glitches, and errors, as long as the predictions are *reasonably* accurate.

It should be clear, from the discussions in Secs. 2.1-2.3, that the accuracy of hydrocode predictions depends on the EOS and constitutive models as much as the numerical methods used in solving the conservation equations. CTH offers many options for these models, and that is one of its strengths. But it also means that one can obtain rather different predictions with different choices for the material models. Hence it is not enough to validate the code. One needs to assess the accuracy of the code when used with *specific* EOS and constitutive models.

The conventional way to evaluate the accuracy of a code is to compare its predictions with experiments. Ideally, one would perform a battery of tests, calculate deviations between theory and experiment, and come up with some kind of a confidence level that accounts for both the experimental and theoretical errors. I am sometimes asked to specify the "error bounds" on my EOS models, the question implicitly assuming that this kind of error analysis can be made.

In most applications involving shock waves, or other high deformations and strain rates, it is not possible to obtain enough experimental data to carry out a meaningful error analysis. In the present study, of hypervelocity dust impacts, there are *no* experimental data in the regimes of interest. It is not possible to com-pute meaningful, quantitative estimates of errors. Statements like "CTH calculations are accurate to at least 20%" are meaningless and should be avoided.

The above observations do not imply, as is sometimes claimed, that all codes and models should be regarded on an equal footing, that one code/theory is as good as another because none of them can be tested. The exact opposite is true.

I will not presume to offer a sure-fire methodology for assessing the accuracy of hydrocode predictions for dust impacts and similar problems. But I will offer a few guidelines that may be helpful.

- One should not assume that the ability to run a calculation means that the results can or should be taken seriously. This principle seems self-evident, but it sometimes gets ignored in practice.

- When developing models, one should be ruthless in testing them against experiment, not try to present them in the best possible light, focusing more on their successes than failures.

- One can, and should, be creative in finding ways to test models, even if the tests are not in the pressure-temperature range of interest. There are several examples of this approach in the present report. See, for example, the sections on the $Al_2O_3$ EOS and the glass impact problems.

- One should perform sensitivity studies using different models. These studies will only be meaningful if the models are significantly different, e.g., one model treats melting and dissociation while the other does not.

- One should support development of improved models and actually use them in the calculations, rather than claim that existing models are good enough just because they have not (yet) been proven to be deficient.

- Model development should be entrusted to those who have shown success in past work—not just in developing theories but getting them to the point where they are actually used in practical problems.

- One should never abandon experiment in place of pure calculation. However, it may be useful to combine calculations with integral experiments to obtain information about material phenomena and properties.

## 2.5  Length-Time Scaling of Hydrodynamic Phenomena

The basic equations of hydrodynamics scale linearly in length and time. This fact is well-known but often ignored when it could be useful in analyzing the results of hydrocode simulations.

Hydrodynamics, in its most basic form, consists of the three conservation laws of physics (mass, momentum, and energy) in continuum form. Consider the Euler form of these equations, written, for simplicity, in one spatial dimension:

$$\partial\rho/\partial t + \rho\,\partial u/\partial x + u\,\partial\rho/\partial x \,=\, 0 \ \text{(mass)}, \tag{1}$$

$$\partial u/\partial t + u\,\partial u/\partial x + \rho^{-1}\partial P/\partial x \,=\, 0 \ \text{(momentum)}, \tag{2}$$

$$\partial E/\partial t + u\,\partial E/\partial x + (P/\rho)\partial u/\partial x \,=\, 0 \ \text{(energy)}, \tag{3}$$

where $x$ is position, $t$ is time, $\rho$ is density, $u$ is particle velocity, $P$ is pressure, and $E$ is internal energy (per unit mass). These equations are to be solved together with the boundary conditions (initial conditions) and the EOS, in the form

$$P \,=\, P(\rho, E). \tag{4}$$

Now consider the linear transformation

$$x' \,=\, ax\,, \ t' \,=\, at\,. \tag{5}$$

It is easy to see that the new equations are identical to the original ones, if the functions $\rho$, $u$, $P$, and $E$ are expressed in terms of the new variables—*provided that the EOS, Eq. (4), is independent of length or time*. Standard EOS and constitutive models satisfy these criteria; important exceptions include reactive materials, which have time-dependent behavior, and materials in which the behavior depends on pore or grain size.

This result has several consequences. First, it means that two hydrocode calculations, differing only in the size scale, will give identical results if all length and time dimensions (including zone sizes and boundaries) are scaled by the same factor (provided the EOS condition is satisfied).

This result also shows why scale-model tests should be expected to give the same results as full-scale tests, again assuming that the EOS condition is satisfied and all dimensions are scaled. Deviations from the scaling law indicate that the material properties may have a built-in time or length scale. Or, in the case of a hydrocode calculation, the zoning and boundary conditions may not be adequate.

These observations should be kept in mind when making hydrocode simulations and in designing experiments to study dust impacts.

## 2.6  Miscellaneous Problem Areas

**Initial State**: In setting up a CTH calculation, it is necessary to specify the initial density and temperature, along with the velocity, of each material region. The default initial state is RTP, the standard condition for the overwhelming majority of hydro problems. The code usually takes care of this problem, and users sometimes forget to specify the inital state when it differs from the default.

In the dust impact problems for SPP, the materials are expected to be at high initial temperatures, 473K for the solar cells, and 1650K for the TPS. If one simply specifies the temperature without giving the density, the code uses the RTP density, and the material is therefore inserted under pressure, giving spurious results. In setting up a problem, one can use the BCAT code [20] to determine the initial density at the initial temperature and pressure desired. I have provided the necessary values for the problems of interest in this project.

**Expansion Into a Vacuum**. The CTH code cannot track the expansion of a gas-like material front into a vacuum. The reason is that, in continuum mechanics, the front of the wave moves with infinite velocity even though it contains no mass. Satisfactory results can usually be obtained by replacing the vacuum with a very dilute gas. I usually recommend air, because that is the most likely gas to be present in terrestrial situations. That choice is less defensible for the SPP, because the solar atmosphere consists primarily of hydrogen and helium. However, it should not make much difference in the present case because the gas is used only to eliminate numerical problems and has no significant effect on the predictions.

**Treatment of Elastic Energy**. CTH contains an error in the treatment of the elastic deformation energy. Since shear deformation contributes to the stress deviators but not to the pressure, the shear deformation energy should be stored separately from the total internal energy and be subtracted from the total when computing the pressure and temperature from the EOS. CTH does not keep track of the two energy terms, thereby predicting a temperature rise in a material undergoing shear. I believe this "thermalization" of the elastic energy is incorrect and have recommended that the code provide a user option to avoid it [21].

Most CTH calculations involve stresses much greater than the yield strength. The code does not provide a way to determine if this effect is really serious. It could be a source of error in some calculations, affecting the accuracy of the strength and fracture predictions in high-strength, brittle materials like $Al_2O_3$ and glass.

**Artificial Viscosity**. CTH, like most finite element hydrocodes, uses artificial viscosity to create shock waves with finite rise times. It is well known that the Rankine-Hugoniot jump conditions still apply in that case, provided the shock velocity is constant. Tests using a Lagrangian code did reproduce those conditions, but tests using CTH did *not* satisfy the energy condition with the default viscosity settings. (The error probably arises in the rezone step.) Reasonable results *can* be obtained by increasing the viscosity settings [22]. But most users do not use the larger values because they smear out the shocks too much. It is not clear if this effect is important in any of the calculations discussed in this report.

## 3. RADIATION CONTRIBUTIONS TO THE EOS

Radiation contributions to the EOS, which are not included in the standard CTH EOS tables, become important at very high temperatures. Since the temperatures encountered in the SPP dust impacts are so much higher than those in typical CTH problems, new EOS tables that include these terms were generated.

The EOSPro code provides an option to include the radiation terms, using the expressions given in Ref. [23]. For each material, the original EOS was read in as a table, the radiation terms were added, and the new table was output on the same density-temperature mesh as the original.

Figure 1 compares the EOS, with and without radiation terms, for fused silica. The pressure is shown as a function of density at 25 temperatures from $10^5$ to $10^8$ K, equally spaced in the logarithm. Isotherms for the EOS without radiation are shown in blue, those for the EOS with radiation in red. The radiation terms are seen to become more important at low densities as well as at high temperatures.

The Hugoniot for fused silica is shown in green in Fig. 1. For a 300 km/s impact of silica on itself, the shock pressure and temperature are 6.3 x $10^4$ GPa and 1.9 x $10^6$ K. A release isentrope from this state is also shown in green.

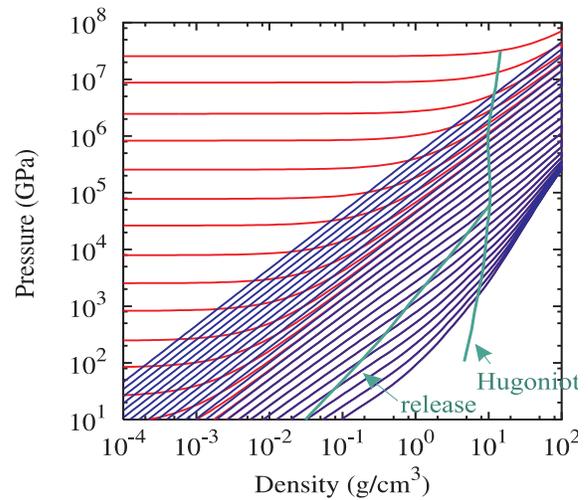

Fig. 1. Effect of radiation on EOS for fused silica. Isotherms range from $10^5$ to $10^8$ K. Blue curves correspond to the EOS without radiation, red with radiation. The Hugoniot is shown in green, along with a release isentrope from the shock state for a 300 km/s impact.

It can be seen that radiation contributions to the pressure are very small for this impact condition. An increase in the impact velocity to about 1500 km/s would be necessary before radiation terms would start to be important. Similar results are obtained for the energy and for the other materials considered in this study.

It appears, therefore, that radiation effects are not likely to play a significant role in the study of dust impacts for the SPP.

# 4.  EOS FOR ALUMINUM OXIDE

Aluminum oxide, $Al_2O_3$, also known as "alumina," is the leading candidate for the top layer of the TPS. This layer will be directly exposed to the sun's radiation and will serve as a radiation reflector and thermal barrier to lower levels of the TPS. This section discusses a new, but preliminary, EOS for $Al_2O_3$.

## 4.1  Existing EOS Models

The CTH database currently offers two EOS tables for alumina [24].[1] The default EOS, material #7412 [25], was calculated using the ANEOS model at Sandia National Laboratories [26][27]; this EOS was used in previous dust impact simulations [3] . The other table, material #7411 [28], was calculated using the GRIZZLY code at Los Alamos National Laboratory [29]. Similar models were used in both cases: the average atom, Grüneisen solid, interpolation to the gas at high temperature and/or low density, and including ionization. These EOS are "global" in the sense that they cover a wide density-temperature range. However, neither model includes a realistic treatment of melting or dissociation.

Figure 2 Compares the Hugoniots calculated from these two models with experimental data (green points), taken from a variety of sources [30]-[34].[2] The upper limit on the data, 2000 GPa, is far short of that expected in a dust impact but high enough to test some important features of the models.

It is immediately clear that EOS 7412, shown by the red line, is too "stiff," especially at the highest pressures, and cannot be regarded as adequate for the present work.

EOS 7411, shown by the blue line, gives reasonable agreement with the data at high pressures and is clearly the better of the two options. It can be regarded as a reasonable starting point for preliminary dust impact calculations. Note, however, that this EOS does not agree with the data in the pressure range 300-750 GPa, probably due to the neglect of melting. It also gives unsatisfactory results for the shock temperature at high pressures, as noted in Ref. [36].

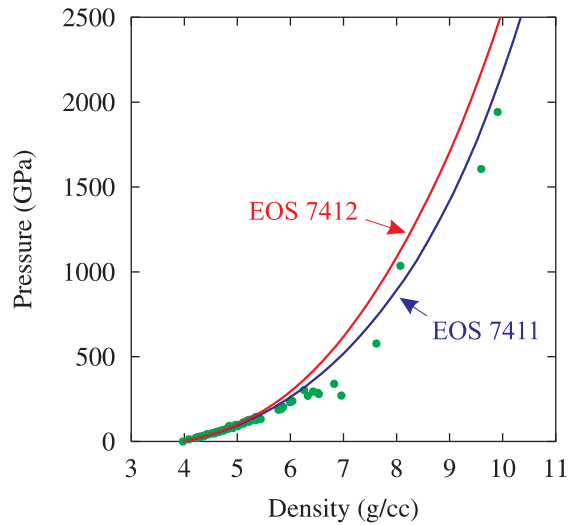

**Fig. 2.  Comparison of EOS models with Hugoniot data for alumina. The green points are experimental data from Refs. [30]-[34]. The solid curves were computed from EOS tables 7411 and 7412 in the CTH database, as discussed in the text.**

---

1.  There is actually a third table, on file seslan, but it is very old and was not even considered for use in the present study.

2.  Unfortunately, EOS data from Ref. [35], which fill in the pressure range 230-1500 GPa, could not be shown in Figs. 2 and 3a because they are not yet published.

Another problem with EOS 7411, as with most tables from the LANL library, is that it does not include the so-called "Maxwell constructions" in the vapor-liquid coexistence region and so does not give a reasonable treatment of vaporization. Fortunately, this difficulty was easily corrected using the EOSPro code. The modified table was provided to APL for use in sensitivity studies.

## 4.2 Limitations of the Existing Model

A more in-depth evaluation of EOS 7411 uncovers two weaknesses, already mentioned above, in the treatment of melting and of dissociation.

Melting in $Al_2O_3$ is expected to occur near 400 GPa and 5000K under shock loading. In a hypervelocity dust impact, most of the material in and immediately around the crater will be in the liquid phase. The properties of $Al_2O_3$ change dramatically on melting, as seen by a large heat of fusion, a large drop in density, and an unusually large increase in the thermal expansion. EOS 7411 does not treat the liquid phase explicitly, relying on a simplistic interpolation scheme. This deficiency may contribute to its inaccuracy at pressures in the range 300-750 GPa.

Another dramatic change in the nature of $Al_2O_3$ occurs at shock states above ~800 GPa and 20,000 K, the onset of dissociation into Al and O atoms. Although EOS 7411 matches the Hugoniot fairly well in the P-ρ plane, it fails to account for the dissociation energy and overestimates the shock temperature by as much as 1.5 x $10^4$ K at very high shock pressures. In a hypervelocity dust impact, the material will be fully dissociated close to the point of impact and much of it will remain dissociated (as well as ionized) as it expands.

## 4.3 Improvements to the EOS Model

The EOSPro code offers all the tools necessary to construct a completely new EOS for $Al_2O_3$, including both melting and dissociation. However, there was not enough time in the present study to complete this task. Therefore, I concentrated on generating a new EOS model that included dissociation, the easier of the two problems, together with a very preliminary and simplistic treatment of melting.

Figure 3 shows the Hugoniot for $Al_2O_3$, in both the P-ρ plane (Fig3a) and T-P plane (Fig3b). Calculations assuming complete dissociation to a mixture of Al + O, shown by solid red lines, were made using the mixture model in EOSPro, together with existing EOS models for aluminum [37] and atomic oxygen [38]. The model predictions agree very well with the experimental P-ρ data for pressures above ~1500 GPa.

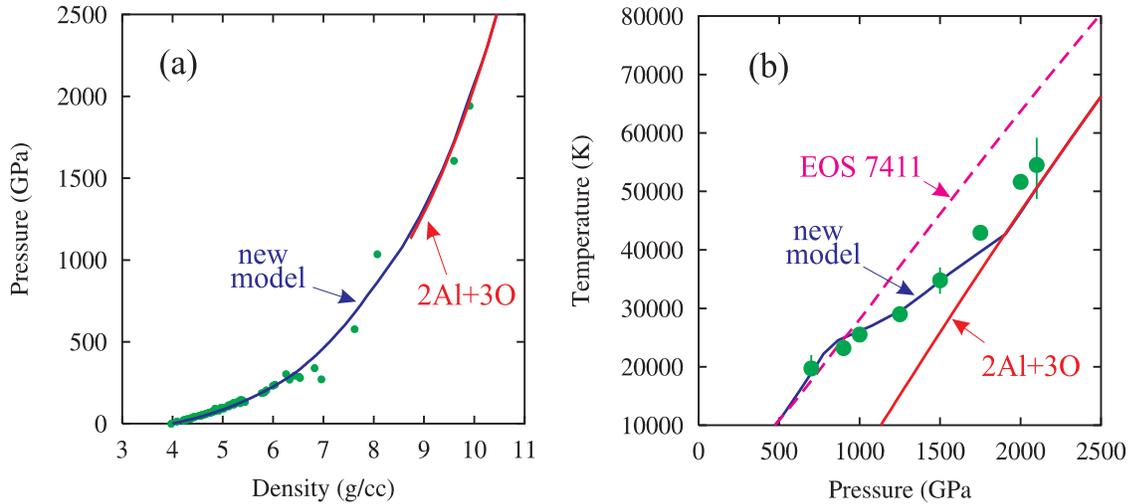

**Fig. 3. Hugoniot calculations for alumina using new EOS model. Experimental data: pressure vs. density (3a) are from [30]-[34], temperature vs. pressure (3b) from [36]. Calculations: red curve—model assuming complete dissociation; blue curve—full model; dashed purple curve—EOS 7411.**

The dissociation model also agrees quite well with the temperature measurements of Miller [36] at high pressures; representative data points, together with error bars, are shown by the green circles in Fig. 3b. By contrast, EOS 7411 overestimates the temperature by up to $1.5 \times 10^4$ K at high pressures, as shown by the dashed purple line in Fig 3b. Miller observed that this discrepancy implies the existence of an "energy sink" that was likely to be dissociation. These calculations confirm his hypothesis.[1]

Examination of Fig. 3 indicates that dissociation begins at about 800 GPa and 20,000K and is complete above ~1900 GPa and 40,000K. Since the effects of ionization are built into the EOS for aluminum and oxygen, the model also accounts for ionization of the mixture. Hence one can have reasonable confidence in the dissociation model to arbitrarily high temperatures.

At the other temperature extreme, one needs a model for solid $Al_2O_3$. This EOS was constructed using a standard average atom/Grüneisen model, matching the model parameters to the low-pressure Hugoniot, thermal expansion, and heat capacity data. The solid model covers shock pressures and temperatures up to the onset of melting at about 400 GPa and 5000K.

---

1. It is important to note that, in the limit of complete dissociation, the calculations do not incorporate any of the experimental data and are completely a priori.

An EOS for the liquid phase is now needed to complete the model. Recent experimental and theoretical work [41]-[44] indicates that there is a significant change in Al-O coordination on melting and that liquid $Al_2O_3$ is ionic. However, it is not yet clear how to model this material. A preliminary attempt, treating the liquid as a mixture of $Al^{+3}$ and $O^{-2}$ ions, did not give satisfactory results. Since a variety of Al-O compounds, e.g., $Al_2O$, $AlO$, and $AlO_2$, are found in the gas phase, it may be that some of the ionic species are molecular and that the composition varies with density and temperature. In that case, one would need a separate EOS for each Al-O compound. The EOSPro mixture model could then be used to construct an EOS for an equilibrium mixture of these compounds, together with the atomic components, Al and O—an extension of the dissociation model.

I had in fact developed preliminary models for the neutral Al-O compounds many years ago, but they are too crude for use in the present situation. That work was never published. Much more work will be needed. However, these old models were used for investigation of the $Al_2O_3$-C reactions, as discussed in Sec. 7.

In the meantime, I generated an EOS for the liquid phase, treating it as a single chemical species and using a simple solid-gas interpolation model, using shifts in the energy and entropy to match the heat and entropy of fusion. This approach gave a rough treatment of melting and was then extended to include the effects of complete dissociation as described above. I will not give details of the model here because I do not intend that it be used for anything beyond sensitivity studies.

The Hugoniot calculated using the full EOS is shown by the solid blue lines in Fig. 3. It agrees fairly well with the measurements in both the P-$\rho$ and P-T planes, although there still appears to be some discrepancy with the data in the range 300-750 GPa. The model also fails to match the large density change on melting [45]. The dissociation transition is clearly evident in Fig. 3b.

This EOS was provided to APL for use in preliminary calculations, especially comparisons with the other EOS tables. I have been told that the new EOS did give significantly different results in the CTH dust impact simulations, but I do have any details to include here.

## 4.4  Constitutive Model

The CTH calculations also require parameters for the von Mises strength model and the minium tensile strength (p-min) fracture model. These properties depend significantly on the type of material. For the present, I have recommended using a shear modulus of 155 GPa, Poisson's ratio of 0.243, and yield strength of 6.3 GPa, values typical of those obtained in experiments by Gust and Royce [46] on various types of alumina. The zero-pressure melting point is 2327 K (0.2 eV in CTH units).

I further recommend a tensile strength of -1.2 GPa, based on measurements of Grady and Wise [47]. Note that these values correspond to room temperature.

Present plans call for the alumina to be vapor-deposited on the second layer of the TPS. The density is reported to be 3.89 g/cm$^3$, compared with a theoretical material density (TMD) of 3.986 g/cm$^3$. This slight porosity should be treated with the p-alpha model, discussed in Sec. 5.2.

It should be obvious that much more work is needed on the constitutive model, along with the EOS. An important complication is that the alumina will be at a very high temperature, 1650K, only 850K below the melting point. The constitutive properties are simply not known at such a high temperature. The solid EOS predicts that the bulk modulus will decrease by about 7.4% at 1600K; I recommend that the shear modulus, yield strength, and tensile strength be decreased by a comparable factor, until better information can be obtained.

The study of dust impacts on glass shows that the tensile strength is an especially important parameter because it controls the extent of spall damage surrounding the central crater. Hence the tensile strength could also be important for $Al_2O_3$. The extent of damage, if very large, could mean that parts of the ceramic could lose their bond to the lower layer and flake off some time after the initial dust impact. This effect may not be as great for $Al_2O_3$ as it is for $SiO_2$, because it has a higher tensile strength. However, further study of this issue is needed.

# 5. POROUS CARBON MATERIALS

Two materials in the TPS for the SPP spacecraft are highly porous forms of carbon—a C-C composite, which has a porosity of about 30%, and a carbon foam, which is 98% porous. It is essential to account for the effects of porosity in the material models for dust impact simulations because pore compaction is an irreversible process that dissipates the energy of a shock wave. Simulations of dust impacts done prior to this study did not include these very important effects.

Porous materials consist of a "matrix," which has a density near to that of the perfect crystal, and void regions (pores), the nature of which depend upon the processes used to create the material. These voids may appear as spaces between grains of the matrix, or they may be more like bubbles. Whatever their origin, the pores compact under the application of pressure. Hence the material model for a porous material will depend on three things—the EOS of the matrix, the pore volume, and the compaction behavior.

Fortunately, models for these two materials can be built upon an excellent EOS for carbon [48] that is reviewed in Sec. 5.1. Section 5.2 discusses the simple but widely-used p-alpha porosity model, in which pore compaction is treated using internal state variables. Sections 5.3 and 5.4 describe CTH calculations in which the microstructures of the composite and foam are treated explicitly.

## 5.1 Carbon EOS

The carbon EOS used in this work is described in Ref. [48]. The model, developed using the Panda code [11], includes three solid phases (graphite, diamond, and metallic) and a fluid phase mixture containing three molecular components ($C_1$, $C_2$, and $C_3$). Separate EOS tables were constructed for each solid phase and for each molecular species in the fluid phase. The mixture/chemical equilibrium model was used to construct a single multicomponent EOS table for the fluid phase from those for the individual chemical species. Finally, the phase diagram and multiphase EOS were determined from the Helmholtz free energies. The model was shown to give good agreement with experimental thermophysical data, static compression data, known phase boundaries, and shock-wave measurements. Thermal electronic excitation and ionization were built into the EOS for $C_1$, so that the model is reliable even at ultra-high temperatures. The final EOS table, which is the default material for carbon in the CTH EOS library, covers a wide range of densities (0 - 100 $g/cm^3$) and temperatures (0 - $1.0x10^8$K).

Figures 4 and 5, taken from Ref. [48], illustrate the importance of a good EOS for the matrix in modeling a porous material.

Figure 4 shows the calculated carbon phase diagram, together with P-T Hugoniot loci for several initial densities in the range 0.12 to 2.16 g/cm³. Figure 5 compares the calculated Hugoniots with experimental data [31][49][50] in the $U_S - u_P$ plane. Shock wave measurements for these initial densities sample the phase diagram over a wide pressure-temperature range. Higher-density materials are shocked into the diamond and metallic solid phases, the transition pressures being dependent upon the initial density. Lower-density materials traverse only the graphite and liquid phases. These differences in the phases are quite visible in some of the Hugoniot measurements.

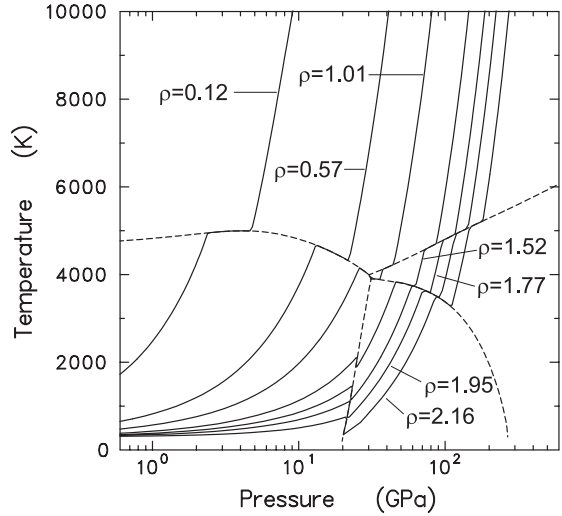

**Fig. 4. Shock-induced phase transitions in carbon. Solid lines are Hugoniots for various initial densities. Dashed lines are calculated phase boundaries.**

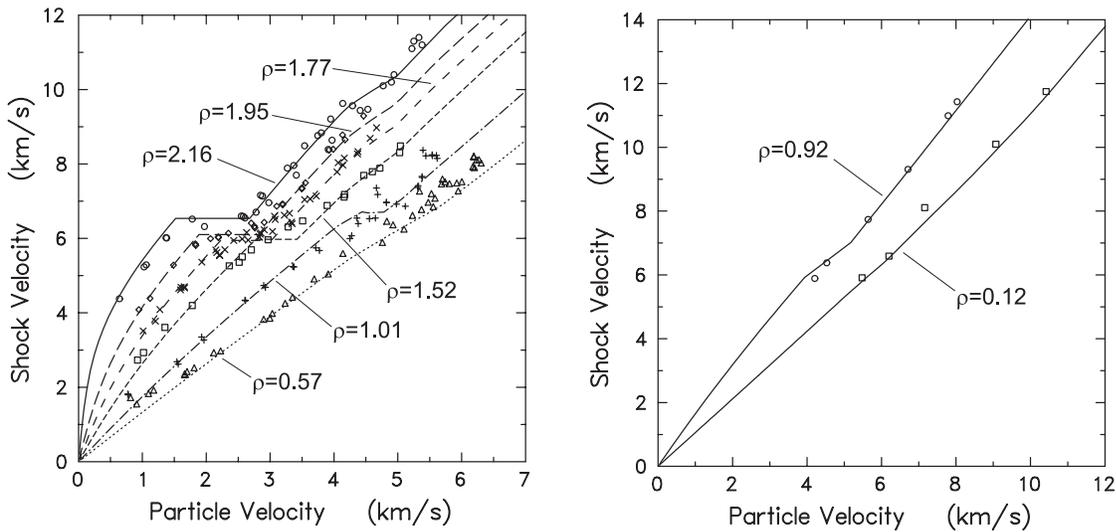

**Fig. 5. Hugoniots for carbon at various initial densities. The solid curves are calculated. Experimental data are taken from Refs. [31][49][50].**

## 5.2 P-Alpha Model

CTH offers the so-called p-alpha model for treatment of pore compaction [9][10]. This model is actually based on ideas taken from work on material response to quasi-static loading conditions and does not treat all of the phenomena that occur in shock compaction. However, it does give reasonable results for some of the ef-fects, including the dissipation of energy by pore compaction. In any case, it is widely used in shock-wave problems, because it is the only practical option available for "production" calculations.

The p-alpha model assumes that the pressure and internal energy of the porous material are given by

$$P(\rho, T, \alpha) = P_M(\alpha\rho, T)/\alpha = P_M(\rho_M, T), \tag{6}$$

$$E(\rho, T, \alpha) = E_M(\alpha\rho, T) = E_M(\rho_M, T), \text{[1]} \tag{7}$$

where $P_M$ and $E_M$ are the pressure and energy of the matrix material, computed from the EOS table, and $\rho_M$ is the density of the matrix with all voids removed. The void fraction is thus described by the function $\alpha = \rho_M/\rho$, an "internal state variable" that is assumed to depend only on the pressure $P$.

The simplest option in CTH uses a quadratic expression,

$$\alpha(P) = 1 + (\alpha_0 - 1)(1 - P/P_S)^2, \tag{8}$$

where $P_S$ is the pressure for complete compaction, i.e., $\alpha = 1$ for $P \geq P_S$. Pore compaction is also irreversible, in that $\alpha$ is only allowed to decrease, i.e., to close up voids but not to open them.

The above option requires only two input parameters—the initial porous density, which determines $\alpha_0$, and the compaction pressure $P_S$. In the absence of experimental data, which is the case in the present study, a "typical" value of $P_S = 0.1$ GPa is usually a good place to start. One should then do sensitivity studies to see if variations in this parameter cause significant differences in the simulations.

In the present study, the calculations thus far, except for those discussed in Secs. 5.3, 5.4, and 6, have been made using the simple model and $P_S = 0.1$ GPa. These calculations gave dramatically different results from the earlier simulations in which porosity was ignored. In particular, the spallation/fracture effects were either reduced or disappeared altogether.

CTH offers additional options for improving on Eq. (8). One can change the exponent in the pressure expression. One can also add an "elastic toe," a low-pressure region in which compaction is reversible. These options have not been employed here because there is no clear way to choose the model parameters. Moreover, it is not clear how much is to be gained by trying to use these refinements. The p-al-pha model has two basic problems that are more fundamental and could lead to even greater errors.

---

1. Here $E$ represents specific internal energy, i.e., energy per unit mass.

The first problem is that the p-alpha model treats the material as spatially homogeneous, the variable $\alpha$ representing an average over the actual grain and pore structure. But spatial variations in the microstructure of the TPS carbon materials are known to be comparable to the dimensions of the impacting dust particles and the thicknesses of the material layers. Sections 5.3, 5.4, and 6 discuss CTH simulations in which this microstructure is treated explicitly, to explore these effects.

The second problem is that the p-alpha model does not accurately treat some of the phenomena that arise in shock-wave compaction. Since $\alpha = 1$ for $P \geq P_S$, the model predicts that the pores collapse instantaneously at high shock pressures. However, mesoscale simulations of shock propagation in porous materials show that pore collapse is not instantaneous and that it broadens the shock front, the thickness typically being 1-2 pore diameters [51]-[55]. It is also found that the kinetic energy generated by pore collapse does not become "thermalized," i.e., converted to internal energy, until well behind the shock front. Hence the state immediately behind the shock front is not an equilbrium state of the material.

The CTH implementation of the p-alpha model treats pore collapse as a time-dependent process [9][10], allowing some experimentation with rate-limited models. One such model was used in Ref. [53] and found to improve agreement with experiments on shock propagation in pumice. An even more elaborate model was developed in Ref. [55]. However, much work on this problem remains to be done, and the rate-dependent models have not yet been incorporated into production versions of the code. Therefore, the effects of these phenomena could not be explored in the dust impact simulations.

There are few data available for estimating constitutive parameters, especially at elevated temperatures. The present study used the von Mises strength model for the composite, with a Poisson's ratio of 0.278, a yield strength of 0.15 GPa, and a melting temperature of 4800K; the tensile strength was taken to be -0.054 GPa, which is a geometric average of the in-plane and interplanar strengths. Because of its low density, the RVC foam was treated as a fluid, and tensile fracture was not allowed.

## 5.3  Carbon-Carbon Composite

Carbon-carbon composites consist of two kinds of carbon—1) carbon fiber bundles woven into cloth, and 2) interstitial carbon that fills the voids between the bundles and also acts as a binder between different cloth layers. The fibers generally have a higher density than the interstitial carbon, leading to a heterogeneous microstructure. Typical dimensions associated with this structure, e.g., the size of and distance between fiber bundles, are hundreds of microns, comparable to and even larger than the dust particles expected in SPP impacts on the TPS.

In order to assess the effects of this microstructure on dust impacts, CTH simulations were made in which the material was explicitly treated as a mixture of two different kinds of carbon. The structure of a carbon-carbon composite is extremely complicated and would be very difficult to model in a CTH simulation; in fact, some details are not known, while others are considered proprietary and difficult to obtain. A complete, detailed model would also require prohibitively expensive three-dimensional calculations. It is not clear that such detail is actually needed to answer the questions. Therefore, a simplified, two-dimensional, model of the structure was developed for the present studies.

The construction of a C-C composite is a complicated process that results in a material having several levels of structure [56]-[60]. For the purposes of this study, the most important facts are as follows.

- The smallest elements of the material are individual carbon fibers with diameters in the range 5-10 μm. They are formed by pyrolyzing polymer strands to drive off other elements (H, N, O). Information about the fiber density is sparse; it is probably close to that of graphite, 2.0-2.2 $g/cm^3$.

- Thousands of carbon fibers are formed into bundles and woven into cloth. Assuming random packing, the voids between fibers *within* a bundle comprise about 17% of the cross-sectional area [59]. These voids will be at least partially filled with the matrix carbon material.

- The cloth consists of two layers of bundles, running in perpendicular directions, interwoven in a variety of possible ways [56][58]. The bundles are roughly elliptical in cross-section, about ten times thicker in the plane than vertical to it.

- A certain thickness of composite is built up by stacking layers of cloth, usually varying the orientation of the weave. This structure is then impregnated with a polymer resin that is allowed to harden, then pyrolyzed to drive off all elements but carbon.

- The impregnating and pyrolysis steps are repeated several times until the desired density is reached. Final densities are typically in the range 1.53-1.60 $g/cm^3$, depending on the number of steps and method used.

In the present study, the composite microstructure was simplified by using only two structural components—fiber bundles and interstitial carbon. The parameters needed for the model are derived in Appendix A. Using a number of reasonable assumptions, the bundles are found to have an elliptical cross-section, 69 μm in thickness and 855 μm in width, spaced 138 μm apart. The densities are found to be 1.85 $g/cm^3$ for the bundles and 1.18 $g/cm^3$, for the interstitial carbon. Hence both components of the structure are themselves porous; the pores within each compo-nent are assumed to be small enough that they can be treated using the p-alpha model.

A complete three-dimensional model of this composite would still be very complicated, even with this two-component model. In order to make a simpler, two-dimensional model, for preliminary hydrocode calculations, it was further assumed that all the bundles were parallel. (Since the problem was run in cylindrical symmetry, they are actually toruses.) The bundles were staggered and arranged in pairs so that there were no paths for an impacting dust particle to channel through the low-density matrix, avoiding the higher-density material.

It might appear that making the bundles parallel, instead of alternating their directions and interweaving adjacent layers, is too unrealistisc a simplification. However, this treatment is actually conservative, for the purposes of the present study, in that it exaggerates the effects of inhomogeneity in the material microstructure; alternating directions and interweaving would increase the homogeneity and cohesiveness of the composite. Therefore, this simpler model is a reasonable place to start the investigation of the effects of microstructure.

The constitutive properties of both components were taken to be the same as those for the homogeneous model, except that the in-plane tensile strength (-0.17 GPa) was used for the bundles and the interplanar tensile strength (-0..55 GPa) was used for the matrix.

CTH calculations using the 2-D composite model are discussed in Sec. 6.

## 5.4  Low-Density Carbon Foam

The third layer of the TPS is reticulated vitreous carbon (RVC), a very low-density foam [61][62]. RVC has an open structure in which thin filaments of vitreous carbon enclose large bubbles. It is formed by polymerization of a resin with foaming agents, followed by carbonization. The result is a material with a high ratio of surface area to volume, minimal reactivity over a wide range of conditions, and low thermal conductivity.

Micrographs show the structure to consist of open cells, or bubbles, averaging 14 facets per cell [61]. The diameters of the openings, or pores, are 50-70% the diameters of the bubbles. Each pore is surrounded by filaments, or struts, that are joined together in a lacy network with tetrahedral connections. RVC is characterized by the number of pores per inch (ppi). 100 ppi corresponds to 50-70 bubbles per inch. The average distance between bubbles is 0.014-0.02 in, about 350-500 μm. Scanning electron microscopy (SEM) of a 100 ppi foam showed struts to have an average length of 130 μm and thickness of 50 μm [62].

RVC is also characterized by its density relative to that of vitreous carbon, 1.65 g/cm$^3$. A 3% RVC would have a density of 0.05 g/cm$^3$.

RVC materials can be made with a range of pore sizes, 5-100 ppi, the bubble diameter decreasing with increasing ppi. The relatively small 425 µm bubbles, typical for 100 ppi, are still larger than expected for impacting dust particles. A 5-ppi RVC would have much larger bubbles, 8500 µm.

In order to assess the effects of this microstructure on dust impacts, CTH simulations were made in which the RVC was treated as high-density carbon with embedded voids. Appendix B discusses both 3-D and 2-D models of the RVC structure. In the 3-D model, the foam is idealized as a body-centered cubic lattice of polyhedral bubbles, with the intersections between these cells representing the struts. This model reproduces the open structure of the foam and gives a reasonable description of the relationship between bubble spacing, strut length, strut thickness, and strut volume or porosity.

As in the case of the composite, Sec. 5.3, full three-dimensional simulations were found to be too expensive, and a much simpler model was devised for use in two-dimensional calculations with cylindrical symmetry. The bubbles were replaced with toruses with a hexagonal cross-section. The most serious deficiency of this model is that it does not reproduce the open structure of the foam, since the sides of the toruses are closed. As a result, it overestimates the strut volume, requiring that the strut density be 0.226 g/cm$^3$, much lower than the value for vitreous carbon, 1.65 g/cm$^3$. Hence the p-alpha model had to be used to treat collapse of the strut material.

Since the struts consist of high-density carbon, it is reasonable to expect that they exhibit some strength. In this work, the constitutive properties were taken to be the same as those of the carbon matrix in the composite, except that the yield strength was dropped by an order of magnitude, to 0.015 GPa.

CTH calculations using the 2-D foam model are discussed in Sec. 6.

# 6. DUST IMPACT ON THE TPS—SAMPLE CALCULATIONS

This section discusses CTH simulations of a dust particle impact on the TPS, using the EOS and constitutive models described in the previous sections. Calculations were made using both the homogeneous and inhomogeneous models of the two porous carbon materials described in Sec. 5.

## 6.1 Description of the Problem

The dust particle was a sphere of fused silica, 100 μm in diameter, with initial velocities of 50 and 100 km/s. The TPS had three layers—0.01 cm (100 μm) of $Al_2O_3$, 0.083 cm of the carbon-carbon composite, and RVC foam extending to the bottom boundary (see below). All materials were inserted at 1600K, the densities adjusted to give zero initial pressure. The room temperature and 1600K densities ($g/cm^3$) were: $SiO_2$—2.204/2.1221; $Al_2O_3$—3.89/3.7536; C-C composite—1.57/1.5134 (average density); RVC foam—0.05/0.048196 (average density).[1]

Figure 6 shows the material regions in the vicinity of the impact particle, at time zero, using the inhomogeneous models of the composite and foam. The sizes and shapes of the components in the carbon materials are discussed in Secs. 5.3 and 5.4 and Appendices A and B.

All EOS tables included radiation terms.[2] The EOS for $SiO_2$ is discussed in Sec. 8; no strength was included, since the entire particle will be melted on impact. The new EOS and constitutive parameters for $Al_2O_3$ are discussed in Sec. 4. The EOS and constitutive parameters for the C-C composite and RVC foam, both homogeneous and inhomogeneous models, are discussed in Sec. 5.

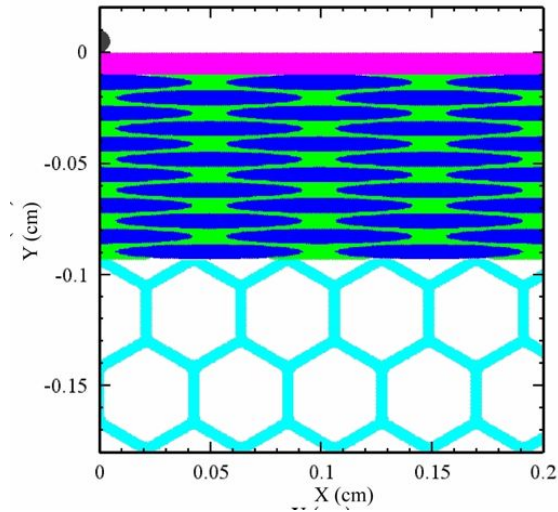

**Fig. 6. Two-dimensional model of TPS dust impact. The $SiO_2$ dust particle is shown in black, $Al_2O_3$ in magenta, fiber bundles in blue, matrix carbon in green, and RVC struts in cyan.**

---

1. For the inhomogeneous models of the composite and foam, a 3.6% reduction in the initial densities of the regions was needed to give the correct inserted mass, i.e., to duplicate the average density of the homogeneous case. The reason for this error is not clear.
2. These terms are not expected to be important, for the reasons given in Sec. 3.

The calculations were run in cylindrical symmetry, the x-axis as the axial direction. The boundaries were $0 \leq x \leq 0.45$, $-0.30 \leq y \leq 0.15$. The zoning package used AMR with nx=ny=32, bx=by=1, and maxl=4, giving a minimum zone size of 8.8 μm (~11 zones across a particle diameter or the $Al_2O_3$ layer). It would of course be desirable to use larger boundaries and finer zones, but that was not possible with my computer system. It is expected that APL will carry out additional calculations.

## 6.2  Results

Calculations for a particle velocity of 50 km/s, run with the homogeneous models of the composite and foam, had just begun to show significant penetration of the composite layer at 2 μs, when the calculation was stopped. Therefore, it was decided to use a higher velocity for testing the inhomogeneous models.

Calculations for a particle velocity of 100 km/s were made using four models of the TPS. Figure 7 shows material plots for all four cases at 2 μs. Figures 8 and 9 show the corresponding density and temperature plots. These figures show only a small fraction of the results from the calculations. Movies of these calculations are also available to help visualize the simulated dust impact events.

Case a is a calculation in which the C-C composite and RVC foam were both treated as homogeneous materials, using the p-alpha model (Sec. 5.2). A hole of radius 0.08-0.09 cm (16-18 times the radius of the dust particle) has been punched through the composite material. A jet of material has penetrated into the RVC foam, having reached a depth of 0.22 cm. This jet consists primarily of low-density, high-temperature material from the composite, along with a few higher-density fragments of spalled material at the outer edges. The composite has been permanently compacted by the shock wave, out to a radius of about 0.14 cm, and shows cracks out to a radius of 0.18 cm. The composite has undergone more compression than the $Al_2O_3$, and the two layers have become delaminated out to a radius of about 0.2 cm.

Case b is a calculation in which the C-C composite is treated as inhomogeneous (Sec. 5.3) and the RVC foam is treated as homogeneous. The results are similar to those for case a, but there are some interesting differences. The radius of the hole is about 25% larger. The tip of the jet is sharper and has penetrated further into the foam. The composite does not exhibit crack formation, and there does not appear to be any delamination with the $Al_2O_3$ layer. These results are probably due to the existence of higher temperatures in the matrix material than those predicted by the homogeneous model of the composite. Mesoscale simulations of shock propagation of mixtures show that softer materials tend to be squeezed between harder materials, forming mini-jets with high temperatures [51][54]. Figures 8b and 9b indicate that the matrix material has vaporized and expanded on the outer regions of the hole in the composite, while the fiber bundles have remained essen-tially intact. Higher temperatures of the matrix material would also explain the sharpness and greater penetration of the jet, since Fig. 7b shows that the tip consists primarily of matrix material.

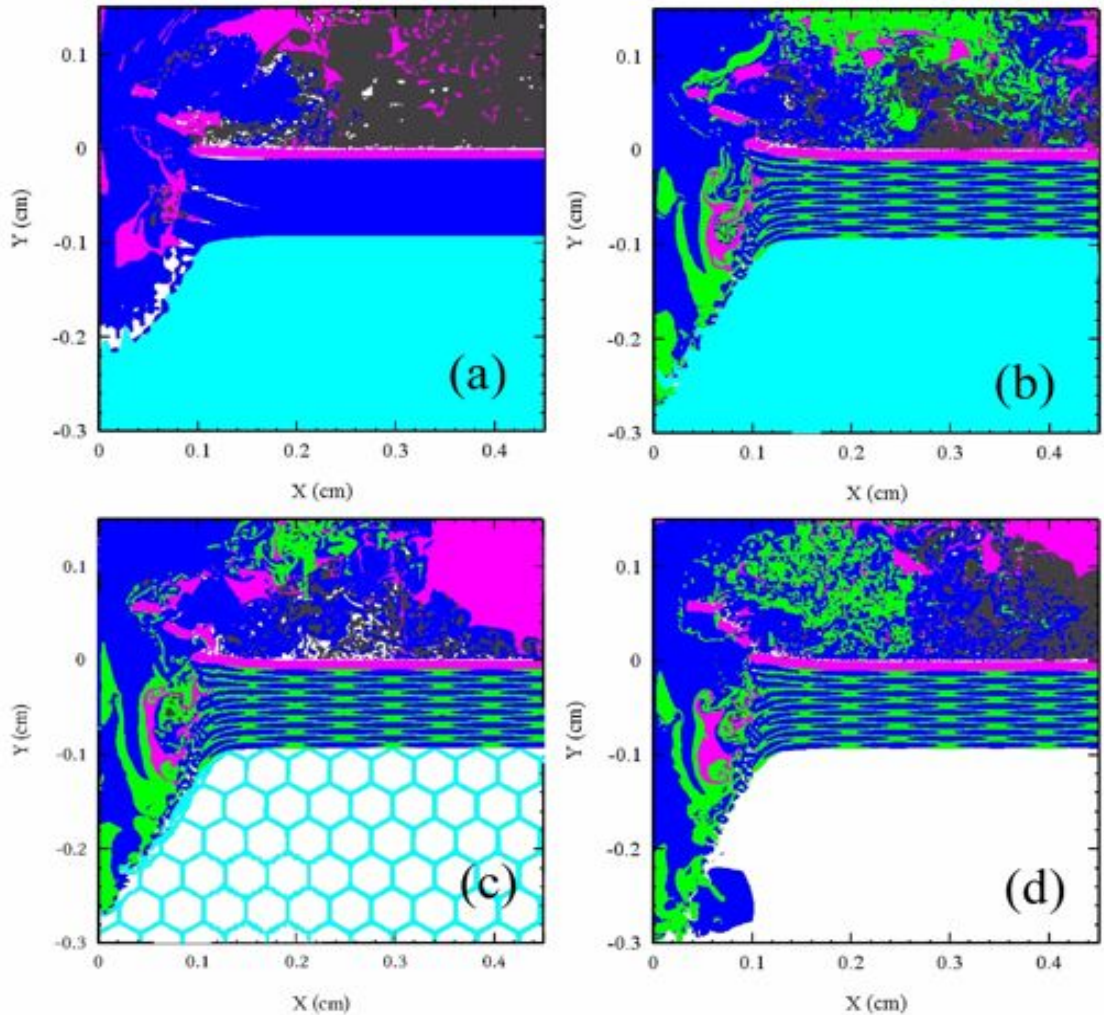

**Fig. 7. Material plots for CTH calculations of a 100 km/s dust impact on the TPS at 2 μs. Case a—homogeneous models of composite and foam; Case b—inhomogeneous model of composite with homogeneous model of foam; Case c—inhomogeneous models for composite and foam; Case d—inhomogeneous model of composite with no foam. Colors represent material regions as in Fig. 6: black—dust particle; magenta—$Al_2O_3$; blue—composite and fiber bundles; green—matrix carbon; cyan—RVC foam and foam struts.**

The higher temperatures could also contribute to the lack of cracks, since the hotter material would have less strength and would not sustain as much shear deformation. The existence of two different material regions in the composite also reduces the development of shear, because the EPDATA set used the MIX=5 option, which does not allow shear in mixed cells containing interfaces.

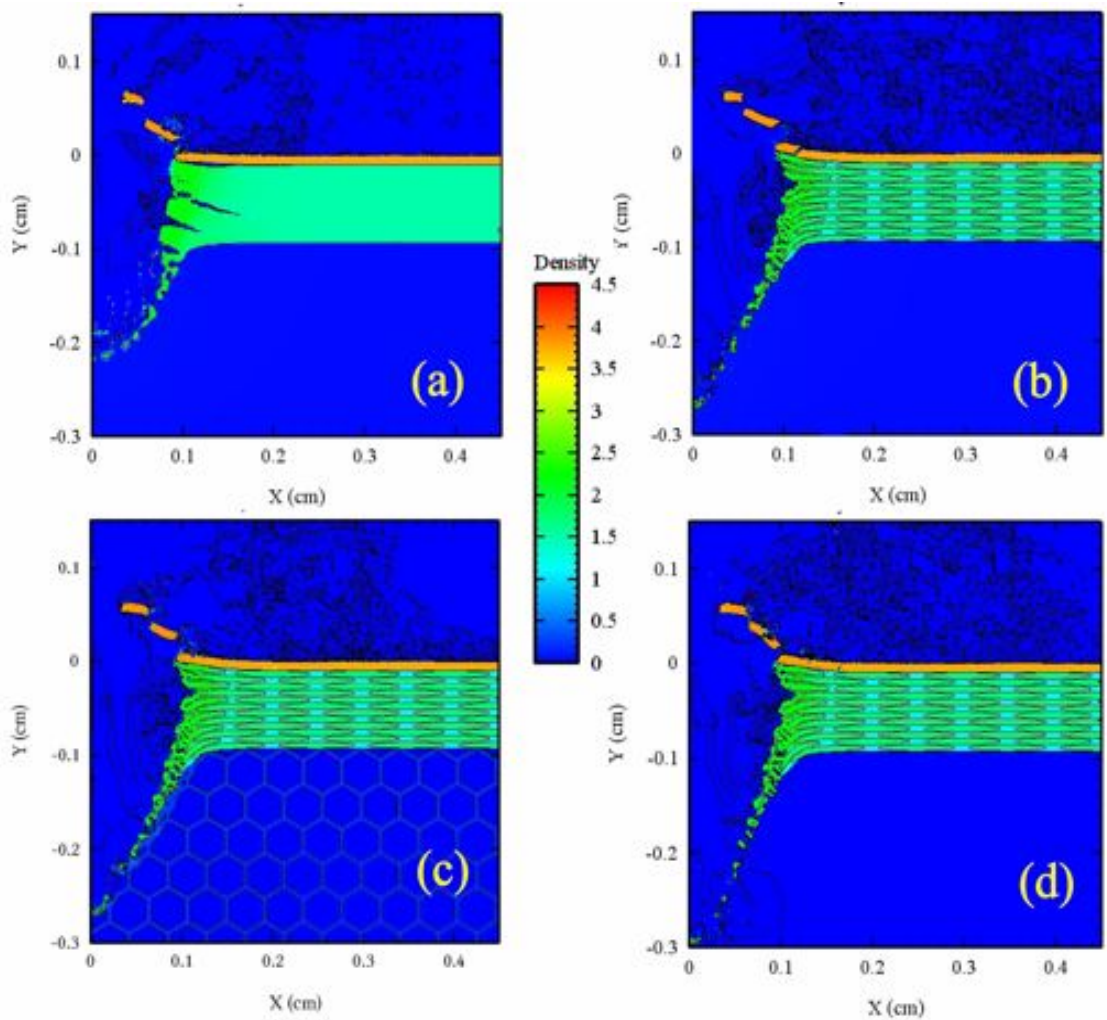

**Fig. 8.   Density plots for CTH calculations of a 100 km/s dust impact on the TPS at 2 μs. The four cases are the same as in Fig. 7. Densities are indicated by color bands, shown in the legend at the center.**

Case c is a calculation in which both the C-C composite and RVC foam are treated as inhomogeneous materials. (This calculation was run at APL [63].) The results look almost identical to those for case b, except for the presence of the hexagons in the undeformed RVC foam. The dimensions and states in the jet are even close to those obtained using the homogeneous foam model. This fact indicates that the foam offers little resistance to the jet and so has little effect on the hydrodynamics of the problem. The jet also causes very little compaction or heating of the foam, at least in this problem. The one subtle difference between cases b and c is that, in the latter problem, the gases in the tip of the jet expand laterally, filling hexagons that have broken open. As noted in Sec. 5.4, the 2-D model fails to reproduce the open cell structure of the RVC foam and so incorrectly inhibits the flow of gases through the pores.

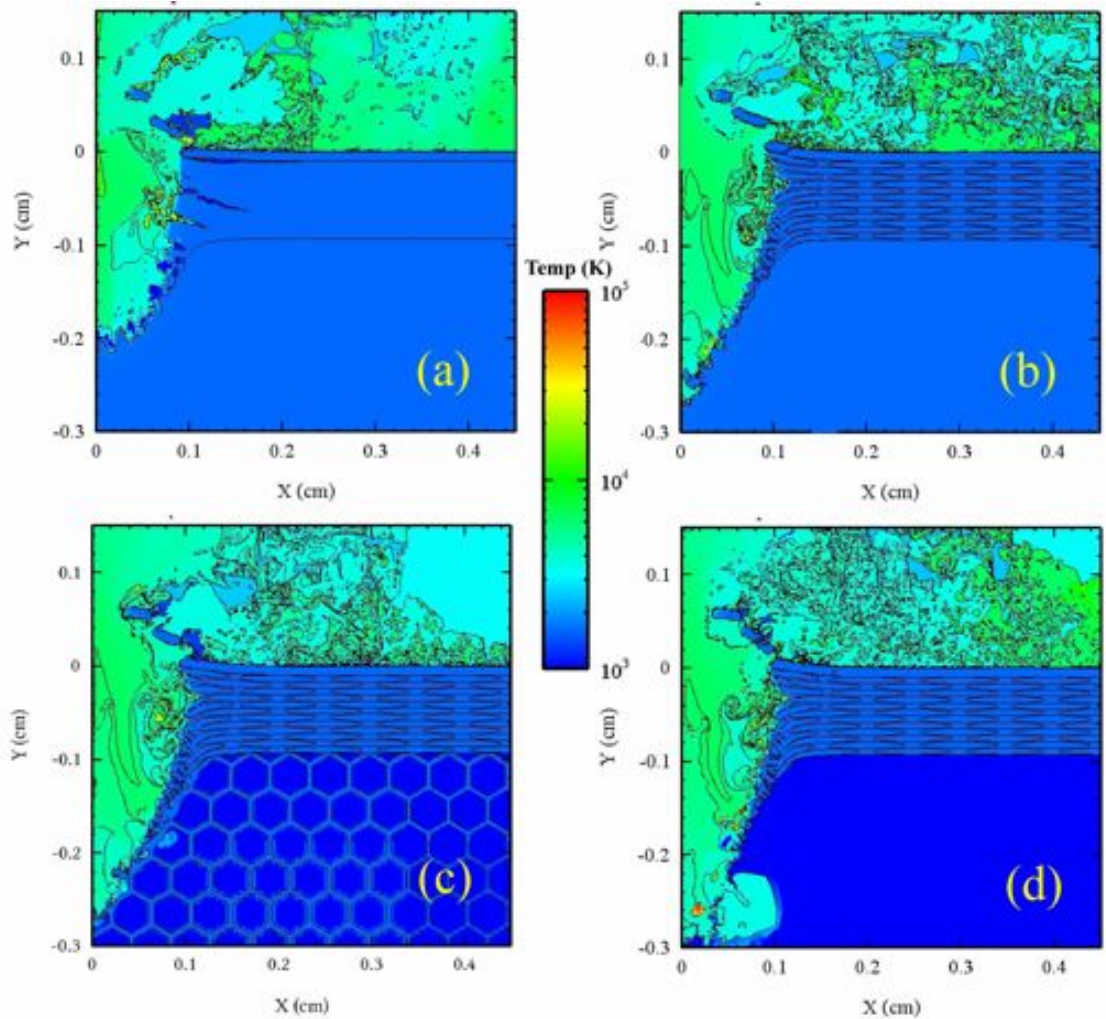

**Fig. 9. Temperature plots for CTH calculations of a 100 km/s dust impact on the TPS at 2 μs. The four cases are the same as in Fig. 7. Temperatures are indicated by color bands, shown in the legend at the center.**

In order to investigate the importance of gas expansion through the open pores of the foam, a fourth calculation was run in which the RVC foam was removed entirely. The results are shown as case d in Figs. 7-9. As expected, case d is virtually identical to cases b and c except near the tip of the jet, where the spalled fragments from the composite have opened up and the jet has begun to expand laterally. This result shows that lateral expansion of the jet could indeed be an important effect, one that requires further study.

Unfortunately, some of the effects seen in case d could be affected by the boundaries, because the jet has just begun to reach the bottom of the computational mesh. Hence this calculation needs to be repeated with larger boundaries and run to much longer times, to investigate the depth of penetration and the extent of lateral expansion. Calculations should also be done using other impact conditions, with larger particles and/or higher impact velocities. If large penetrations and expansions are observed, it will be necessary to estimate how much these low-density gases can actually damage the structure of the foam.

The main results of this section can be summarized as follows:

- The microstructure of the composite has several effects on the predictions—the size of the hole, the nature of the jet, the formation of cracks around the hole, and delamination between the composite and the $Al_2O_3$ layer. These effects, however, are probably minor when it comes to making engineering decisions about the SPP. A more elaborate 3-D model of the composite could be made, but it is not immediately obvious whether or not it is necessary.

- The RVC foam offers little resistance to the jet formed by impact of the dust particle on the top two layers of the TPS. However, the open pore structure does allow the jet gases to expand laterally as well as penetrate more deeply. Additional 2-D calculations should be made to scope the extent of this expansion. If large expansions are observed, a model will be needed to estimate how much damage will be done to the foam structure. 3-D calculations can also be made, but it is not yet obvious that they are necessary.

# 7. REACTIONS BETWEEN CARBON AND ALUMINA

The oxygen in $Al_2O_3$ can oxidize carbon to form either CO or $CO_2$, leaving behind Al-O suboxides or elemental Al. Some of the possible reactions are listed below, along with the heats of reaction at 298K.[1]

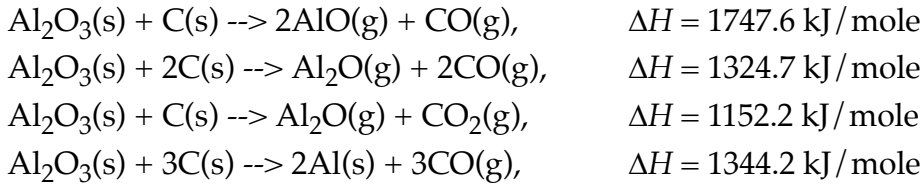

$$Al_2O_3(s) + C(s) \longrightarrow 2AlO(g) + CO(g), \qquad \Delta H = 1747.6 \text{ kJ/mole}$$

$$Al_2O_3(s) + 2C(s) \longrightarrow Al_2O(g) + 2CO(g), \qquad \Delta H = 1324.7 \text{ kJ/mole}$$

$$Al_2O_3(s) + C(s) \longrightarrow Al_2O(g) + CO_2(g), \qquad \Delta H = 1152.2 \text{ kJ/mole}$$

$$Al_2O_3(s) + 3C(s) \longrightarrow 2Al(s) + 3CO(g), \qquad \Delta H = 1344.2 \text{ kJ/mole}$$

Because these reactions are all endothermic, they only occur at temperatures high enough for entropy effects to offset energy effects. Reaction between $Al_2O_3$ and carbon has been reported to be the initial step in formation of AlN, when nitrogen is present, in the temperature range 1648-1825K [65]. These studies also found that $Al_2O$ is more abundant than AlO, in the temperature range studied, and that CO is much more abundant than $CO_2$.

Reactions between alumina and carbon during a dust impact could have an effect on the crater formation, because conversion of the reactants to more volatile products ($Al_2O$, Al,O, CO) would lead to higher pressures at a given density and temperature. This effect would not be important at very high temperatures, where CO is dissociated and is not involved in the reactions.

Chemical equilibrium calculations were made to investigate the importance of these reactions on the EOS of a mixture of $Al_2O_3$ and carbon. These studies used the mixture model in the EOSPro code, together with EOS tables for a variety of chemical species: solid $Al_2O_3$, solid and liquid Al, atomic and molecular oxygen, CO, $CO_2$, and four Al-O compounds—AlO, $Al_2O$, $AlO_2$, and $Al_2O_2$. EOS tables for the four Al-O compounds are very preliminary, having been developed during previous unpublished work on aluminized explosives. However, they do offer a starting point for the present study.

Figure 10 shows the effects of density and temperature on the reaction between alumina and carbon, for an equimolar mixture of the two materials. The calculated concentrations of carbon (red curves) and CO (blue curves) are plotted as functions of temperature at nine densities ranging from 0.001 to 4.0 g/cm$^3$. (This mixture would have a density of about 3.7 g/cm$^3$ at RTP.) The concentration of $CO_2$ is not shown, being almost negligible over the entire range.

---

1. Heats of reaction are calculated from heats of formation for the reactants and products [64]. A positive heat of reaction indicates that the products have a higher enthalpy than the reactants, so that the reaction absorbs enthalpy/energy from the surroundings.

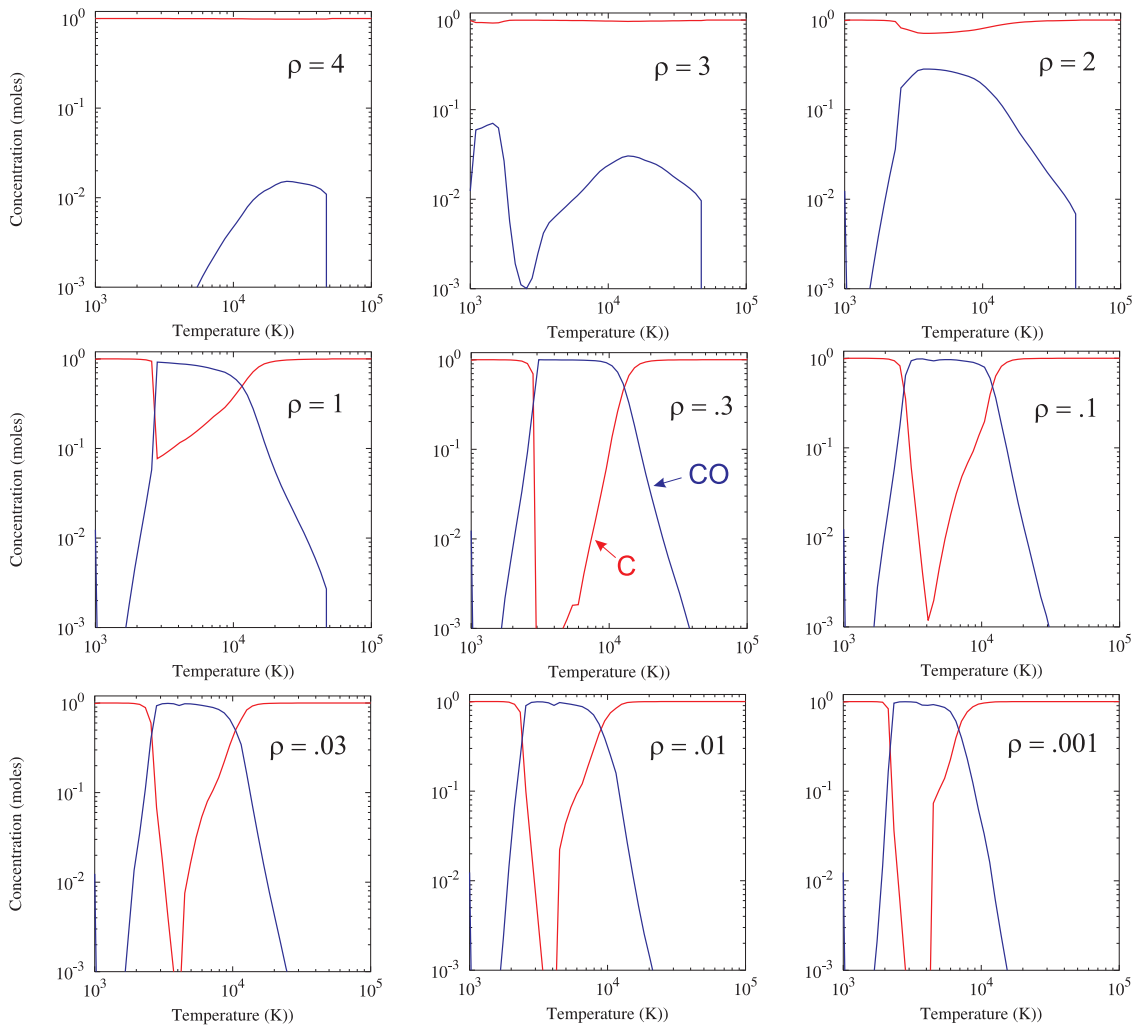

**Fig. 10. Alumina-carbon reaction as function of density and temperature. Results are shown for an equimolar mixture, Al₂O₃ + C. Red curves show concentrations of carbon as functions of temperature at the densities indicated. Blue curves show concentrations of CO.**

Reactions between alumina and carbon are shown by a drop in carbon concentration and a corresponding increase in CO. Figure 10 shows that there is very little reaction at the two highest densities, $\rho = 3.0$ and $\rho = 4.0$, the carbon concentration being almost independent of temperature. For $\rho = 2.0$, the calculations show a 20% drop in carbon over the temperature range 2000-15,000K, and this drop becomes more pronounced at $\rho = 1.0$. The amount of reaction becomes very significant at the lower densities.

These calculations indicate that reactions occur at temperatures in the range 2000-15,000K. As noted previously, reactions do not occur at low temperatures because they are endothermic, driven by entropy. They are not observed at very high tem-peratures because CO dissociates and does not contribute to the equilibrium. The calculations predict that $Al_2O$ is the primary product at the lower temperatures, AlO becoming more important at higher temperatures, and Al dominating at still higher temperatures.

Figure 11 shows the effects of reaction on the EOS of an equimolar mixture of alumina and carbon. Pressure is shown as a function of energy, rather than temperature, because the dependence on internal energy is much more important than the dependence on temperature in hydrodynamic flow. The red curves show the pressure vs. energy at ten densities, as indicated. For comparison, the blue curves show the results for a nonreactive mixture, in which CO and $CO_2$ are not allowed to form.

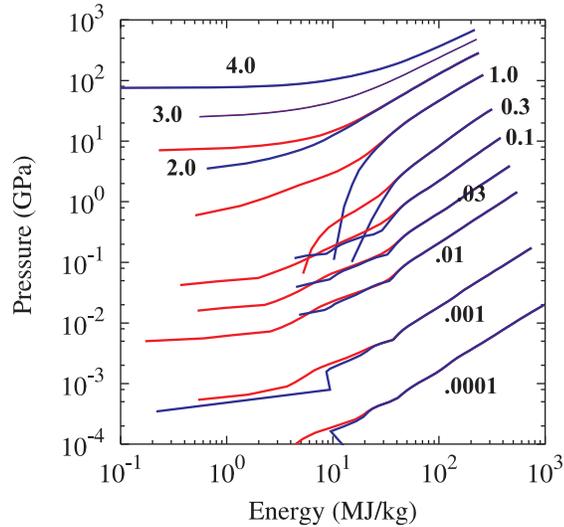

**Fig. 11. Effect of reaction on EOS of an alumina-carbon mixture. Red curves show pressure as function of internal energy for reactive mixture at densities indicated. Blue curves show results for non-reactive mixture.**

These results show that the reactions can have a significant effect on the EOS, but only when the temperature is below about 15,000K and the density of the mixture is in the range 0.3-2.0 $g/cm^3$. In such cases, formation of volatile CO and Al-O compounds leads to higher pressures than in the unreacted mixture. Reactions do not have a significant effect on the EOS at densities above 3.0. Even though reactions do occur at low densities, as seen in Fig. 10, they have only minor effects on the EOS; the reason for this result may be that all of the mixture components are volatile in this regime.

Examination of the calculations for the 100 km/s impact on the TPS, discussed in Sec. 6, indicate that $Al_2O_3$-carbon reactions are *not* likely to have a significant effect on dust impact phenomena.[1] In the early stages of the impact, vaporized $Al_2O_3$ fills the crater as it begins to develop but does not mix with the carbon. As more carbon is produced, it pushes most of the $Al_2O_3$ out of the crater, so that most of the mixing between the two components occurs in the plume, where the densities are very low and the temperatures are very high. Reactions could occur in the plume but would not be expected to influence the hydrodynamic phenomena.

---

1. The observations discussed here are best made by examination of movies of the CTH simulation results.

A small amount of $Al_2O_3$ does remain inside the crater, mixing with the carbon, as can be seen in Fig. 7. The temperatures inside the crater do fall within the critical range discussed above, i.e., below 15,000K. However, there is a very sharp jump in density at the boundary between the shocked, compacted carbon and the gases inside the crater, so that there is very little material in the critical density range 0.3-2.0 g/cc. Hence most of the density-temperature states seen in the simulations are outside the range where reactions are expected to have an effect.

Examination of the 50 km/s impact also leads to much the same conclusions. In that case, there was more mixing between $Al_2O_3$ and carbon *inside* the crater. Hoever, the existence of a sharp jump in density at the boundary of the crater eliminates most density states in the critical region.

Perhaps it would be prudent to make further studies. CTH simulations could be made at other impact conditions. But the most important step would be to improve the EOS tables used for the Al-O compounds, as noted above. The reaction calculations should be repeated after those improvements have been made.

The CTH code does not have the capability to treat reactions among different materials in the EOS input set, even in mixed cells, because it does not allow diffusion of one material into another.[1] This capability would be useful in other problems besides the present one. One example is the modeling of secondary reactions of explosive reaction products in air, an effect that is believed to be important for such explosives as TNT [66].

Unfortunately, this capability would be very difficult to install as a standard option in CTH because it would require major changes to the structure of the code. If it is decided to pursue further investigations of this issue, I recommend writing a separate post-processing program to analyze the mesh variables, searching for density-temperature states where reactions could occur and estimating the effects on pressure and temperature.[2] If that analysis, together with improved EOS models for the Al-O compounds, showed that reactions really could be important, the next step would be to explore ways to construct a special version of the code tailored to this specific problem.

---

1. I am not aware of any other code that has this capability.
2. Programs of this type were used in analysis of mesoscale simulations, as discussed in Refs. [51][52][54][55].

# 8. EOS FOR GLASS

## 8.1 Overview

The SPP solar cells are expected to consist of a 20-mil (0.058-cm) thickness of glass, covering thinner layers of adhesive and germanium on a metallic substrate. These cells could experience dust impacts with velocities up to 300 km/s, even higher than those for the TPS. A reliable EOS and constitutive model for glass are essential for accurate numerical simulations of these impacts. Fused silica is also being used to model the dust particles in impacts on the TPS and solar cells.

The cover glass for the solar cells is expected to be some type of cerium-loaded microsheet. Such materials typically contain 59-69% $SiO_2$, along with other oxides, including $ZnO$, $B_2O_3$, $CeO_2$, $Al_2O_3$, $TiO_2$, and $Na_2O$ [67]. The composition has not yet been specified, and the properties of the material are not yet known.

The present investigation has been limited to fused silica, $SiO_2$, the simplest type of glass and the one that has been studied most extensively. The CTH EOS database already offers a sophisticated EOS for this material [68]. That model provides a good starting point for the present study, although there are areas where more work is needed. This EOS is reviewed in Sec. 8.2 and its limitations are noted.

By contrast, there is no good model for the shear and tensile properties of fused silica. This deficiency is quite serious, because the constitutive properties have a profound effect on the results of dust impacts, particularly the size of the fracture-damaged region outside the central crater [17][18][69]. Therefore, a significant fraction of the time under this contract has been devoted to this problem.

Section 8.3 discusses a constitutive model for fused silica that includes the von Mises model for the shear behavior, with a realistic formula for the melting curve, and a p-min tensile strength. The tensile strength parameter is then fixed by matching CTH calculations to experimental data from low-velocity impact tests. Those calculations are discussed in Sec. 8.4. Section 8.5 summarizes the main findings and the work still to be done.

## 8.2 Fused Silica EOS

There are two common allotropic forms of silicon dioxide—alpha quartz, a crystalline material with an ambient density of 2.648 g/cm$^3$, and fused quartz, also known as fused silica, which has a random atomic structure and an ambient density of 2.204. Alpha quartz is more common in natural geological materials, including sand grains. Fused quartz is formed by rapid cooling of the liquid and is actually metastable with respect to alpha quartz.

Both allotropes transform to the stishovite phase, which has an ambient density of 4.287, under pressure. The transition has a profound effect on the Hugoniot for both materials, but the transition process deviates from equilibrium and is not yet understood. The melting behavior of the two materials is quite different. Both materials dissociate, to atomic silicon and oxygen, at high temperatures. Temperature measurements at high shock pressures show evidence of an energy sink due to dissociation [70], like that seen in $Al_2O_3$ (Sec. 4).

EOS models for alpha and fused quartz are described in Ref. [68]. Both models, developed using the Panda code [11], include three phases—the ambient phase, stishovite, and the fluid, which includes dissociation and ionization at high temperature. Separate EOS tables were constructed for the solid phases, for liquid $SiO_2$, and for atomic Si and O. The mixture/chemical equilibrium model was used to construct a single multicomponent EOS table for the fluid phase from those for $SiO_2$, Si, and O. The phase diagram and multiphase EOS were determined from the Helmholtz free energies. The EOS tables in the CTH EOS library, cover a wide range of densities (0 - 100 g/cm³) and temperatures (0 - 1.0x10⁸K).

Figures 12 and 13, taken from Ref. [68], show the phase diagram for fused quartz and the Hugoniot in the $P$-$T$, $U_S$-$u_P$, and stress-density planes. The model predicts that fused quartz will transform to the stishovite phase at about 8 GPa, with a striking 80% increase in density. The calculations are in good agreement with the experimental data for the high-density phase. Predictions for the low-density phase are complicated by strength effects, as discussed below. Questions remain about the detailed behavior in the transition region.

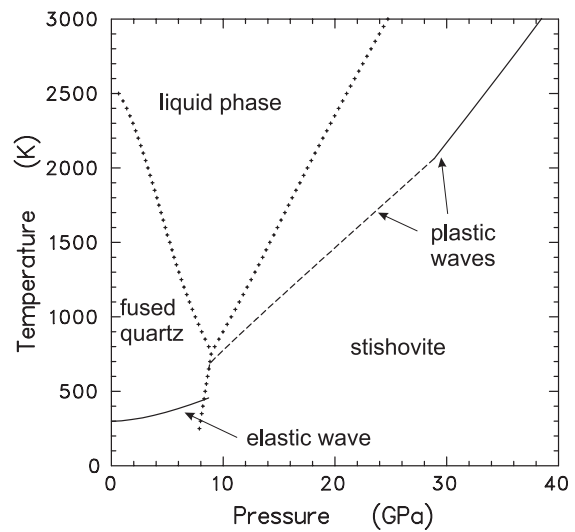

**Fig. 12. Phase diagram and Hugoniot for fused quartz. The phase boundaries are shown by crosses. The Hugoniot locus is shown by solid and dashed curves.**

Figure 12 also shows a pronounced drop in the melting temperature of fused quartz with increasing pressure. The Hugoniot temperature in stishovite is just below the fused quartz-stishovite-liquid triple point. Reference [68] noted that the model overestimates the zero-pressure melting point by 560C and possibly also the triple point. Hence formation of the liquid phase could play a role in the shock-induced transition. Further studies of the solid-solid transition and melting behavior are needed.

Reference [68] also showed that dissociation has a significant effect on the quartz Hugoniot at shock pressures above 200 GPa; the model gave good agreement with shock data from Ref. [73] when dissociation was included. Shock temperature measurements [70], taken several years later, showed that the model did reproduce the energy sink due to dissociation at high shock pressures, although there were discrepancies in the range of the transition.[1]

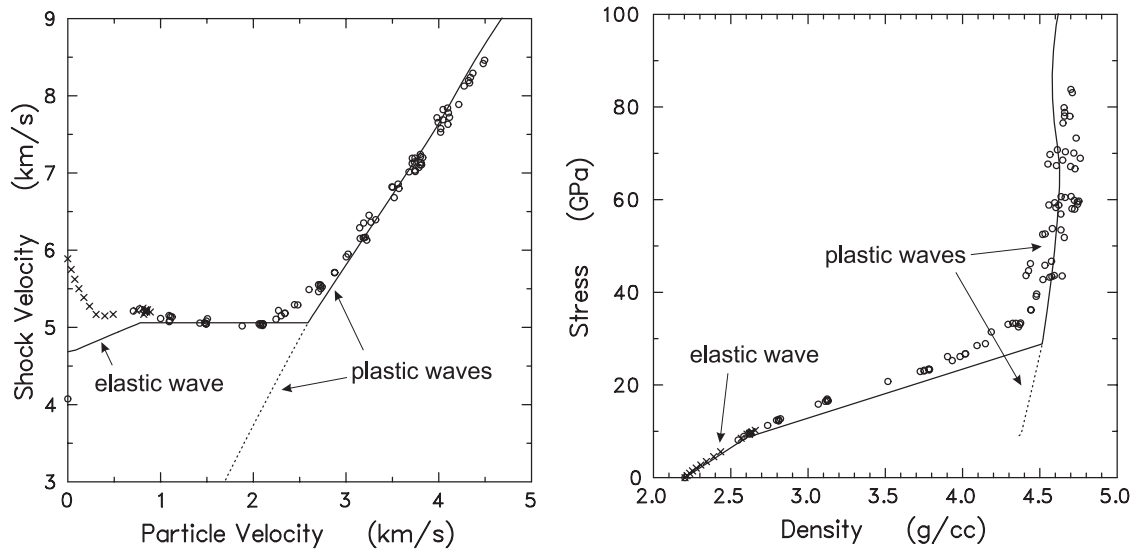

**Fig. 13. Hugoniot for fused quartz. Experimental data [31][71][72] are shown as circles and x's. Curves are calculated using the multiphase model of Ref. [68].**

The principal goal of the work described in Ref. [68] was to develop EOS for composite materials, including silica phenolic, for which the quartz EOS was needed. Therefore, the report did not make a detailed comparison with all of the available data, and it noted that certain aspects of the quartz model should be regarded as incomplete and preliminary. More work is especially needed on the stishovite transition, the liquid model and melting, and the treatment of dissociation.

## 8.3 Constitutive Model for Fused Silica

Hydrocode simulations of dust impacts on fused silica show that the constitutive properties play a central role in the predicted results, especially in the nature and extent of spall damage around the central crater. A satisfactory model of these properties does not exist at present.

The shear behavior of fused silica exhibits two features that are somewhat anomalous. The first is the nonlinear behavior of the shock velocity in the elastic regime as seen in the x's in Fig. 13 [71]. The marked drop in $U_S$, for $0 \le u_P \le 0.5$ km/s, corresponds to a decrease in the shear modulus from about 35.4 GPa at zero pressure to about 13.6 GPa at the Hugoniot elastic limit (HEL). The strength model used in Ref. [68] does not reproduce this non-linear behavior. The calculated curve in Fig. 13 used a shear modulus of 13.6 GPa and yield strength of 4.2 GPa for the ambient phase, these values chosen to match the HEL reported in [74]. The strength was set to zero above the HEL.

---

1. The experiments of Refs. [70] and [73] were made on alpha-quartz. However, the dissociation effects are equally important for fused quartz because the two materials have identical EOS in the stishovite and liquid phases.

The second unusual feature is that fused silica loses all of its strength after yielding [75]. The HEL also appears to coincide with the onset of the transition to stishovite, at least to within experimental error. Reference [68] speculated that melting could be the reason for this loss of strength, that improvements to the liquid model at low pressures could lower the melting temperature so that the elastic wave terminates at the melting boundary. If so, that could explain the fact that fused quartz undergoes permanent densification to a new amorphous phase near 10 GPa, in both dynamic and static experiments [75][76].

In the present study, no attempt has been made to model the nonlinear behavior because sensitivity studies showed that the dust impact predictions were far more sensitive to the tensile strength than to the shear modulus and yield strength. Therefore, all the calculations discussed here were made using the SGL model [15][16] with a shear modulus of 28.6 GPa; this value is midway between the maximum and minimum values given above and is also close to that used in the hydrocode simulations of Davison, et al. [17][18]. The yield strength was taken to be 4.2 GPa, as in [68]. The SGL melting temperature formula was fit to the calculated melting curve of Ref. [68]; the result was

$$T_{melt} = T_0 \exp[2A(1 - 1/\eta)]\eta^{2(\gamma - A - 1/3)}, \ \eta = \rho/\rho_0, \tag{9}$$

where $T_0 = 2500$ K, $A = 9.8$, $\gamma = -1.1$, and $\rho_0 = 2.204$ g/cm$^3$.

The model also included temperature-dependence of the shear modulus and yield strength. CTH parameter BST was set to 1.1, assuming that the temperature-dependence of the strength parameters is the same as that of the bulk modulus, as computed from the EOS. This term will have only a minor effect on the results.

All of the CTH calculations discussed in this report employ the simple p-min spall model, as discussed in Sec. 2.3. Each material in the calculation is assigned a tensile strength parameter PFRAC, which is negative. This parameter can be used with either a pressure-driven or stress-driven fracture criterion. If the tensile pressure/stress in a given computational cell drops below the specified value, void is added to the cell and the material density is increased to bring the pressure/stress back to zero. Sensitivity studies of dust impacts showed that the stress criterion gave more realistic predictions of spallation behavior in glass than the pressure criterion. The stress criterion was then used in subsequent calculations.

An additional user-defined parameter, PFVOID, specifies the tensile stress for any cell that already contains void; the tensile stress then becomes the higher value of the two (negative) parameters PFRAC and PFVOID. Sensitivity calculations showed that realistic behavior could only be obtained if cells containing void were easily fractured. The calculations reported here used PFVOID = -1.0E5 dynes/cm$^2$ (-0.1 bar), which is essentially zero.[1]

With all other parameters thus fixed, the challenge is to choose the tensile strength for glass. The first set of calculations were made with a value of -0.04 GPa, the incipient spall strength for quartzite,[2] given by Davison and Graham [77]. This value predicted far too much spall damage when used in CTH simulations of the glass impact experiments discussed in Sec. 8.4, below. A large number of sensitivity studies were then made, varying not only the tensile strength and other parameter settings, discussed above, but also examining the effects of zoning and boundary conditions. The best results to date have been obtained using a tensile strength of -0.60 GPa.

## 8.4 CTH Simulations of Glass Impact Tests

There are no experimental studies of dust impacts on glass at impact velocities and conditions relevant to the SPP project. However, there are a number of observations at lower velocities, involving meteoroids and space debris impacting mirrors and solar cells, including laboratory tests. This work has been discussed by Davison, et al. [17][18] and by Drolshagen [69]. Davison, et al., also discuss numerical simulations of impacts using the AUTODYN hydrocode.

It is well-known that impacts on brittle materials, like glass, generate a central crater that is surrounded by a larger region of spall damage. The central crater is formed when the impacting particle penetrates into the target, displacing and ejecting material. The outer damaged region forms later, as rarefactions produce tensile states leading to damage and spall. Numerical simulations of dust impacts on solar cells require a material model that can reproduce four key measurements—the crater diameter and depth, and the spall diameter and depth.

Analysis of the problem indicates that the crater dimensions are determined primarily by the EOS and shear behavior of the material, the relative importance of the two terms depending on impact velocity. The EOS would be expected to become more important as the impact velocity is increased and more material in the vicinity of the impact is melted. Drolshagen states that the nature of the material ejected from the crater depends markedly on impact velocity. For velocities up to 5 km/s, the ejecta consists of solid fragments. For velocites above 20-25 km/s, it is vaporized. For velocities in the range 5-20 km/s, it consists of solid fragments, liquid material, and vapor. For SPP dust impacts, involving velocities ≥ 50 km/s, the crater dimensions should be determined more by the EOS than the constitutive properties. Therefore, impact experiments at low velocities do not test those aspects of the model that determine the crater size in the regime of interest to SPP.

---

1. Sensitivity calculations showed that the predictions depend on the value of PFVOID. Hence it is necessary to fix this parameter while varying the value of PFRAC. The sensitivity to PFVOID may decrease with decreased zone size, as the problem approaches numerical convergence. However, this issue has not yet been investigated.
2. Quartzite is a geological material containing both alpha quartz and fused quartz.

By contrast, the dimensions of the spalled region are determined primarily by the tensile behavior of the material at the low stresses produced in the outer regions of the impact. Hence the same tensile model should apply to all impact velocities. While experiments at the higher velocities relevant to SPP would be most useful, CTH simulations of the lower-velocity tests are also worthwhile for the purposes of this study. They offer a way to calibrate the tensile strength parameters and to test the ability of the code to predict certain trends, especially the dependence of the crater and spall dimensions on impact velocity.

Unfortunately, examination of the experimental data uncovers certain problems. In order to simulate a particular test, one needs to specify the size and velocity of the impacting particle. This kind of information is not readily available for the impacts observed in samples recovered from space. Questions also arise about the type of glass used in the tests, about statistical variations in the results, and about differences between different equations that have been developed to fit the data. There has not been time, during the present preliminary investigation, to review all of the available data and identify what experiments would be most useful for testing the model and simulations.

The present study has focused on simulations of two sets of impact tests made at Auburn University (AU) and reported at the 2003 Hypervelocity Impact Symposium by Davison, et al. [17][18]. The AU tests used fused quartz particles impacting glass mirrors at impact velocities of 6.2 and 9.9 km/s, the reported results being averages over three separate experiments. These two sets of tests were made using different size impactors, 62 μm for the low-velocity tests and 124 μm for the high-velocity tests. It was decided to use the same size particle, 100 μm, for the CTH simulations, which scale in length and time. The results of the AU tests, when scaled to a 100 μm particle, are listed below:

|  | 6.2 km/s | 9.9 km/s |
|---|---|---|
| crater radius | 74 μm | 27 μm |
| crater depth | 156 μm | 60 μm |
| spall radius | 550 μm | 208 μm |
| spall depth | 82 μm | 27 μm |

The above results are clearly inconsistent, in that all four dimensions are much smaller for the high-velocity test than the low-velocity test. This result does not agree with the simulations reported in [17], with the trends observed in other work, or with the CTH simulations done in this investigation. Davison, et al., state that the discrepancy is due to the fact that different types of glass were used in the two sets of tests. The higher-velocity tests were made on ultra-low expansion (ULE) glass, a material that is usually made by adding $TiO_2$ to pure $SiO_2$. The lower-velocity results are an average of two tests on fused silica and one on ULE.

It is reasonable that differences in composition or method of fabrication could have a significant effect on the strength of the glass and thus on the spall dimensions. It is not clear that those differences would also have such a dramatic effect on the crater dimensions. If so, this effect would have significant implications for the design of SPP solar cells. However, examination of the three lower-velocity tests, made by P. K. Swaminathan [63], showed that the ULE results were not dramatically different from the fused silica results at 6.2 km/s. Hence there remains a serious inconsistency in the ULE glass results at the two velocities.

Faced with these observations, it was decided to ignore the data for the higher velocity AU test, at least for the present, and concentrate on finding a tensile strength that would give reasonable agreement with the lower velocity tests.[1]

Figure 14 shows snapshots of the CTH calculations for a 6.2 km/s impact, using a tensile strength of -0.6 GPa and the other model parameters discussed above. 2DC geometry was used as in the TPS simulations (Sec. 6). The boundaries were $0 \leq x \leq 0.16$, $-0.10 \leq y \leq 0.06$. The zoning package used AMR with nx=ny=32, bx=by=1, and maxl=4, giving a minimum zone size of 3.1 μm (~32 zones across a particle diameter). The effects of zoning and boundaries are discussed below.

Figure 14 shows that the effects of the impact occur in three main steps. In the first step, penetration of the particle into the target displaces material and forms a crater; the behavior in this step is controlled primarily by the EOS. The second step involves spallation and ejection of material around the top of the crater, widening the crater diameter at the top but not the bottom. The last step is spallation damage at a still wider radius, the development of a crack that starts beneath the surface and propagates to the top. The resulting damaged material is not ejected but remains in place in this simulation. The dimensions of the crater and spall region are in reasonable agreement with the experimental values given above.

---

1. The information about the individual low-velocity tests was obtained after all of the analyses reported here had been made. Therefore, the model calibrations were made to match the average results of all three tests, including the one with ULE. However, the differences between the fused silica and ULE are not significant enough to warrant redoing all the computations at this time.

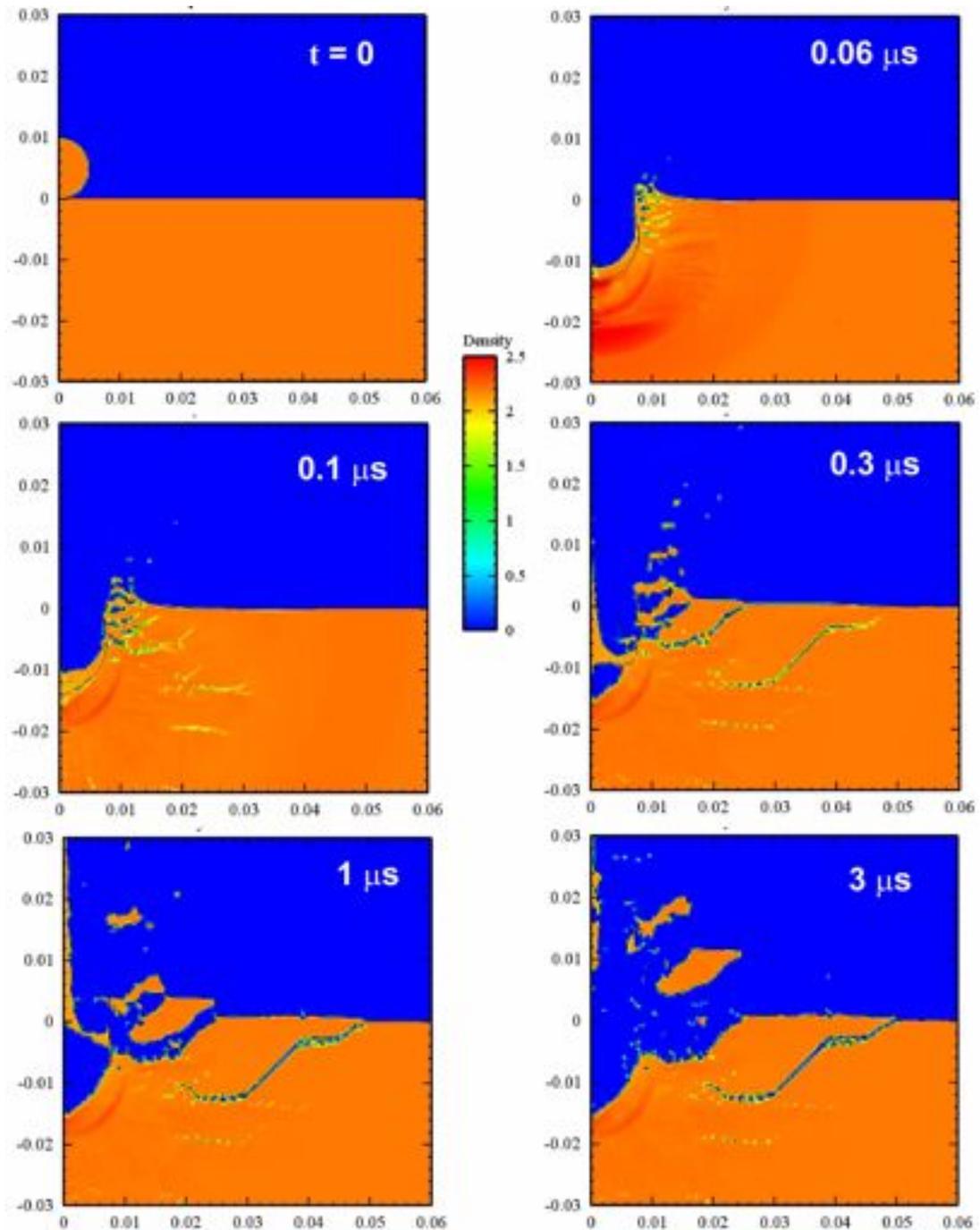

**Fig. 14. CTH calculation of a particle impact on glass. A 100 μm sphere of fused silica impacts a semi-infinite fused silica target at 6.2 km/s. Each snapshot shows the density, using the colors in the central legend; blue indicates void, red a density of 2.5 g/cm$^3$ or higher. Three phenomena are observed: crater formation by displacement of target material, crater enlargement by fracture and ejection of material, and damage of the target out to a distance of about four times the crater diameter. Calculation was made in 2-D, with cylindrical symmetry, using the EOS and constitutive models discussed in the text.**

Figure 14 shows a ring of glass around the crater that is more dense than the rest of the target. Examination of the calculation shows that residual motion from the impact is the cause of this compression. The phenomenon still persists at 3 µs but should damp out at later times. It is not a permanent densification of the material.

Although the results discussed above are encouraging, further examination of the simulations reveals that they are not numerically converged. Figure 15 compares snapshots from three CTH simulations, at 0.7 µs. These calculations used the same EOS and constitutive models, differing only in the boundaries and zoning.

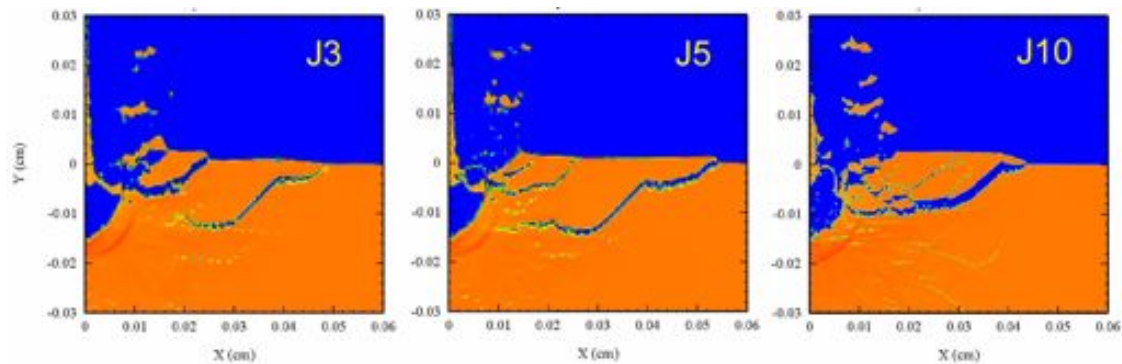

**Fig. 15. Effects of zoning and boundaries on glass impact calculations. Snapshots correspond to 0.7 ms. Run J3 is identical to that shown in Fig. 14. Run J5 has double the boundaries but the same zone size as J3. Run J10 has the same boundaries as J5 but finer zoning.**

Snapshot J3 is taken from the calculation discussed above and shown in Fig. 14. Snapshot J5 was carried out with twice the boundary size but the same zone size ($0 \leq x \leq 0.32$, $-0.22 \leq y \leq 0.10$, nx=ny=32, bx=by=2, and maxl=4.) The results for J5 are similar to those for J3, but it predicts a larger spall region.

Snapshot J10 used the same boundaries as J5 but smaller zones. Since the AMR setup used in these calculations used the presence of interfaces to refine the zoning, the target in J10 was inserted as two separate but identical materials, spheres of fused silica embedded in a fused silica matrix. The parameter maxl was also increased to 5.[1] The resulting decrease in zone size had a dramatic effect on the predictions, reducing the size of the spall region and extending the large crack all the way to the crater, allowing the damaged material to be ejected.

The sensitivity of the calculations to zone size is particularly disturbing. Additional calculations have been made at APL, using maxl=6 and maxl=7. The calcu-lations do not show signs of converging, the size of the spall region increasing with maxl=6, then decreasing again with maxl=7. Further work on this issue is needed, as discussed below.

---

1. The EPDATA set was also run with MIX=1, to ensure that shear forces could develop across the material interfaces in the target.

Before concluding this section, I will explain why I claim the new EOS/constitutive model discussed in this report to be superior to that of Davison, et al. [17][18]. First, their EOS does treat the stishovite transition and also includes hysteretic behavior, but it does not include any temperature dependence; this omission is probably not justified even for the low-velocity AU tests, let alone for the much higher impact velocities important to the SPP dust impacts, where the material will be melted and even dissociated. Their shear strength model is more elaborate than the one used here, but it is not clear that the features included have a significant effect on the predictions. In any case, their model does not predict the nonlinear behavior of the shear modulus. It does include loss of strength above the HEL but does not model melting, as done here. Finally, their model also uses a p-min tensile criterion, just as is done here. These differences may be the reason that their values for the shear and tensile strength are much different than those used in this work.

## 8.5  Work to be Done

The model development discussed in this section represents a significant step forward. A great deal has been accomplished in a rather short period of time. However, there is much more that needs to be done. I will conclude this section by noting some of the issues that need to be addressed.

**Numerical convergence**. Every effort should be made to achieve numerical convergence, so that one can have confidence in the predictions. There are still a number of approaches that could be tried, but this work will require considerable time and patience—and computing power. If it proves impossible to achieve full convergence, it will be necessary to ensure that the same zone size is used in all computations, when making comparisons between different impact conditions.

**Improvements to the EOS**. As already noted above, improvements to the EOS model are needed in the modeling the non-equilibrium and hysteretic behavior of the stishovite transition, melting and properties of the liquid phase, and the dissociation transition. It will also be necessary to make many more comparisons of the model predictions with experimental data, including wave-profile measurements [74][75] as well as impact tests.

**Improvements to the constitutive model**. The non-linear behavior of the shear modulus should be added to the model. (This feature can probably be treated by modifying the SGL model [16].) Coupling of shear and tensile damage should also be added using the JFRAC model [19] as a starting point. Other improve-ments may also be needed. But all changes should be evaluated by careful comparison with wave-profile experiments and other data.

**Comparisons with experimental data**. All model improvements, including the EOS and constitutive properties, need to be tested against as many kinds of experiments as possible. An extensive review of the literature will be needed to determine what data already exist.

One of the major problems encountered in the present study was the lack of accurate, reliable test data on dust impacts. It may be necessary to design new experiments in which all relevant parameters are carefully controlled and measured.

**Effects of composition**. As noted above, the glass used in the SPP solar cells will be 59-69% silica, along with other oxides, including $ZnO$, $B_2O_3$, $CeO_2$, $Al_2O_3$, $TiO_2$, and $Na_2O$. The EOS and constitutive properties of the real coverglass could be quite different from those of fused silica. The tools for developing an EOS for this material exist, but it will take time to apply them to this problem. Development of a constitutive model will require new data—shock-wave experiments, not just measurements of quasi-static properties and impact tests. In fact, it may be possible to design a type of glass that will be more resistent to damage than ordinary silica.

The time and effort needed to gain an understanding of these effects is likely to be a major challenge to making reliable numerical simulations of dust impacts.

**Additional CTH calculations**. Most of the CTH simulations done so far have been for impacts on semi-infinite targets at low velocities. Many more calculations are needed to examine the effects of velocity and finite-thickness on the predictions. Of course, the usefulness of these calculations will depend on the solution to numerical convergence and model calibration problems.

I have already set up a suite of eight calculations, varying the velocity from 6.2 to 300 km/s, to see how the crater and spall behavior change with velocity. I believe that the nature of spall damage could change significantly at the higher velocities and that the ratio of the spall diameter to the crater diameter may not increase with velocity as has been observed in past calculations. Since the crater and spall dimensions increase with impact velocity, it was necessary to increase the size of the boundaries, while keeping the zone size fixed, at the higher velocities. My computer system is not powerful enough to run the problems at the higher velocities, and I have sent them to APL to be run.

# 9. EOS FOR SILICONE ELASTOMER

## 9.1 Introduction

The second layer of the SPP solar cells will be a 3-mil adhesive bond between the coverglass and germanium layers. The adhesive of choice, DC93-500, is made by Dow Corning specifically for use in solar cells [78]. It is a transparent silicone elastomer, formed by the polymer poly(dimethylsiloxane) (PDMS), together with cross-linkers and (possibly) fillers.

The composition of DC93-500 appears to be proprietary and unavailable in the open literature, and there are no EOS data for this material. Fortunately, data do exist for a similar material, Sylgard 184 [79], which is also made by Dow Corning and consists of the same ingredients, albeit in different proportions. Sylgard also has nearly the same density as DC93-500 (1.05 vs. 1.08). Some details of the Sylgard composition are also proprietary, but enough is known to construct a model.

The CTH EOS database already offers an EOS table for Sylgard, material 7931 [80], but it is based on a simplistic model that was deemed inadequate. The adhesive layer is thin, but the material is much softer than either the coverglass or the germanium, and an accurate treatment of its properties could be important for predictions of dust impacts. Therefore, a new EOS model was developed using the EOSPro code, employing methods previously applied to models of polymers.

Like most polymeric materials, Sylgard is a reactive material. The polymeric form is metastable at low temperatures but decomposes when heated. Decomposition occurs in the range 20-30 GPa on the Hugoniot. An EOS model for this type of material consists of three main parts: 1—an EOS for the unreacted material, 2—an EOS for the reaction products, and 3—a model for the dissociation reaction.

## 9.2 Unreacted Polymer

An EOS for the unreacted polymer was created using the EOSPro solid-gas interpolation option, fit to sound speed, thermal expansion, and heat capacity data [80][81], and Hugoniot data up to 25 GPa [31][82]. The model combines an empirical solid model with a transition from solid-like lattice vibrations of the nuclei to gas-like degrees of freedom at low densities and/or high temperatures. (See Sec. 4 of the Panda manual [11] for details.) In the present work, the gas-like modifications were used to obtain realistic (but only qualitative) treatment of the vaporization behavior of the polymer in the low-density/low-temperature regime.

The solid model is usually employed with the average atom approximation, in which there is one atom per unit cell and all lattice vibrations are characterized by a single Debye temperature. That approach overestimates the heat capacity for polymers and other materials in which many of the lattice vibrations are high-frequency internal modes. In the present case, satisfactory results for the heat capacity were obtained with three atoms and six internal vibrations per unit cell. The frequencies of the internal modes were taken from Dowell [80].

The model requires specification of a Grüneisen parameter, which was originally chosen to match the experimental thermal expansion coefficient. However, that gave an unreasonably low density, 0.5 g/cm$^3$, for the polymer at 473K, the working temperature for the SPP solar cells. The 473K density was estimated to be about 0.89 g/cm$^3$, by assuming the thermal expansion coefficient to be constant, and the Grüneisen parameter was then chosen to match that value. However, there are no experimental data to support the validity of this estimate.

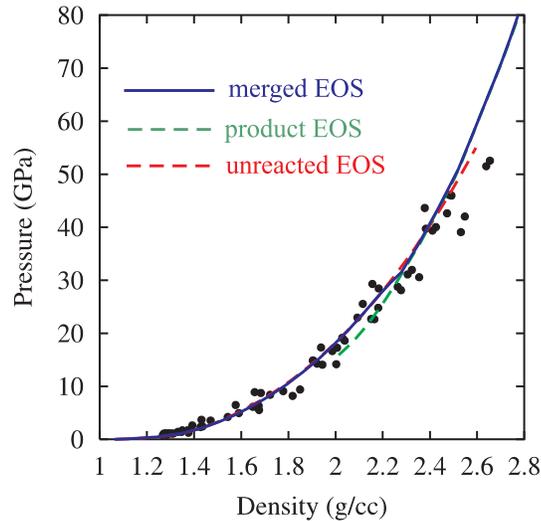

**Fig. 16. Hugoniot for Sylgard. Circles are experimental data [31][82]. Solid blue curve calculated from merged EOS table. Dashed red and green curves calculated from unreacted & reaction product tables.**

Figure 16 compares the Hugoniot computed using the EOS for the unreacted polymer (dashed red curve) with the experimental data.

It should be noted that a 15% expansion of the adhesive at 473K could have significant effects on the structure of and distribution of stress within a solar cell. Measurements of the adhesive properties at high temperatures are needed.

## 9.3  Reaction Products

The mixture-chemical equilibrium model in the EOSPro code [11] was used for the reaction product EOS. [This model was also used in treating the dissociation of Al$_2$O$_3$ (Sec. 4) and glass (Sec. 8), and for the study of Al$_2$O$_3$-carbon reactions (Sec. 6).] An EOS for the reaction products is already available in the CTH database (material 7943), calculated with the same model used here. A new EOS table was created because the older one did not go to a high enough temperature and did not agree well with newly-discovered Hugoniot data [82].

The mixture-chemical equilbrium model requires three things: the elemental composition, the heat of formation of the unreacted material, and EOS tables for all chemical species that can be formed in the reaction products.

The repeat unit for the basic polymer, PDMS, is -$C_2H_6SiO$-. The actual elastomer composition depends on the cross-linker compounds and any fillers used. The elemental composition of DC93-500 and Sylgard are apparently proprietary. The present work used the Sylgard composition given by Dowell [80], which was obtained by chemical analysis at Los Alamos National Laboratory: 30.9% C, 7.4% H, 37.8% Si, 23.9% O by weight, or C[2]H[5.708]Si[1.046]O[1.161].

A surprisingly large range of values have been reported for the heat of formation of Sylgard. The value used in this work, -7.782 MJ/kg, was taken from a Purdue University website [83]. The source of this datum, which falls in the middle of the reported values, is not given, and it is not obvious how to evaluate its accuracy. Fortunately, this value gives satisfactory agreement with the Hugoniot data, and a 30% change in this parameter would not have much effect on the predictions.

Eighteen compounds were included in the reaction product equilibrium. These included twelve CHO species—$CO_2$, $H_2O$, CO, $CH_4$, $O_2$, $H_2$, O, H, graphite, diamond, and liquid carbon—for which EOS tables are available from previous work. (The EOS for liquid carbon included $C_1$, $C_2$, and $C_3$, as also noted in Sec. 5.) Five Si compounds were included—alpha quartz, coesite, stishovite, liquid $SiO_2$, and atomic Si, the EOS tables taken from [68] and [85]. Solid SiC was also included, using the EOS table given in Ref. [84].

Figure 16 compares the Hugoniot computed using the EOS for the reaction products (dashed green curve) with the experimental data..

## 9.4 Merged EOS Table

The best way to treat the dissociation reaction would be to use a reaction rate model, as has been done in EOS for other polymers [51][68][86][87]. However, the existing experimental data are not sufficient to determine the exact dissociation pressure, let alone come up with a model for the rate parameters. Since time was also limited, it was decided to use the simpler approach of merging the two EOS into a single table. The unreacted EOS was used below 2500K, the product EOS above 2500K. Since there is a small drop in the internal energy upon reaction, the temperature points in the mesh from 1900 to 2800K were omitted, in order to avoid a negative heat capacity, leaving the EOS in this region to be determined by interpolation. This solution is not perfect, but it does match the Hugoniot, gives the correct behavior below and above the transition, and allows the solar cell impact calculations to get underway. The EOS number of the merged table is 7935.

Figure 16 compares the Hugoniot computed using the merged EOS table (solid blue curve) with the experimental data. The merged table follows the unreacted EOS up to about 29 GPa, the reaction product EOS above ~32 GPa, with a small increase in density near 30 GPa.

# 10. GERMANIUM EOS

The third layer of the SPP solar cells will be a 6-mil thickness of germanium, which is obviously a critical component. No EOS table for this material was found when searching available databases and the internet. Therefore, the EOSPro code was used to construct an EOS table, using an approach previously used for silicon [85]. The model described here is fairly sophisticated, in that it includes two solid phases, the liquid phase, and thermal electronic excitation. However, it should be regarded as preliminary because so little time was available to work on it; several issues need further attention, as noted below.

Ge has the cubic diamond structure at RTP but transforms to a number of higher-density phases under pressure [88]. The most important high-density phase, which has the beta-tin structure, is 21% more dense [89]. The transition pressure is about 10 GPa. Transitions to other phases occur at higher pressures, but the density increases are much smaller, about 2%, and it was decided not to include these phases in this preliminary model.

Ge melts at 1211K, at zero pressure, with an increase in density. The melting temperature drops with pressure until it reaches a triple point at the solid-solid phase boundary [90]. The melting temperature then increases with pressure above the triple point. This behavior is somewhat unusual but not really anomalous, being seen in a number of other materials—including water, silicon, and fused quartz.

The nature of the Ge melting curve has two important consequences. First, the effects of melting on the EOS surface cannot be taken into account using simplistic melting models, like the Lindemann formula that is often used in EOS codes. Second, shock-wave propagation induces melting at a rather low pressure, so that the liquid phase becomes the only important phase at shock pressures above ~40 GPa.

The EOSPro phase transition model was used to create the EOS table. Separate EOS tables were first constructed for the diamond, beta-tin, and liquid phases. Highlights of these models are given below. The model parameters are given in the EOSPro input files, which are listed in Appendix C.

- The EOS for the cubic diamond phase, Ge-I, used the Birch-Murnaghan formula for the zero-Kelvin (0K) isotherm and the Einstein model for the lattice vibrational terms. Parameters were fit to the ambient density and sound speed, thermal expansion coefficient, and heat capacity. No thermal electronic term was included because Ge-I is an insulator.

- The EOS for the β-Sn phase, Ge-II, used the Birch-Murnaghan formula for the 0K isotherm and the Debye model for the lattice vibrations. The 0K parameters were chosen to match the room temperature isotherm (up to 75 GPa) and the Hugoniot data. Parameters for the Debye model were chosen to give a I-II phase boundary in accordance with experiment [90]. The Thomas-Fermi-Dirac model was used for the thermal electronic excitations.

- The liquid EOS was constructed using the CRIS model, which is based on statistical-mechanical perturbation theory, using the solid 0K isotherm to compute the interatomic forces [91]. The 0K isotherm parameters were taken to be the same as for Ge-II. Parameters that define the 0K isotherm in tension, as well as some parameters in the liquid model, were chosen to match the melting and boiling point properties. A statistical weight of 4 was also added to match the entropy at the melting point. The TFD model was also used for the thermal electronic terms.

The EOSPro phase transition tools were used to compute the phase boundaries and construct the multiphase table from the EOS tables for the individual phases. The input file is given in Appendix C. The calculated phase diagram, shown in Fig. 17, is in good agreement with the experimental data [90]. The calculated Hugoniot is shown in blue. Shock melting begins at about 22 GPa and is complete at about 41 GPa; all shock states above this pressure correspond to the liquid phase.

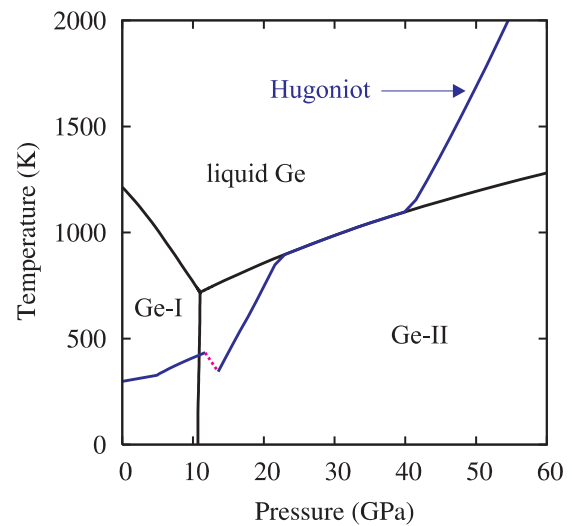

**Fig. 17. Phase diagram and Hugoniot for germanium. Phase boundaries are shown in black, the Hugoniot locus in blue. The magenta section connects the elastic and first plastic wave with the second wave.**

Figure 18 compares the calculated Hugoniot for Ge with experimental data [31][71][92][93] in the $U_S$-$u_P$ and stress-density planes. The Hugoniot for Ge exhibits a three-wave structure up to a pressure of about 35 GPa. The first wave is the elastic precursor, the second corresponds to the onset of the I-II transition, and the third corresponds to shock states above this transition. This third wave is shown in magenta in Fig. 18. At stresses above 35 GPa, the third wave overtakes the second wave and there is a two-wave, elastic-plastic structure. At pressures above ~60 GPa, the plastic wave overtakes the elastic precursor and there is only one wave.

Gust and Royce have made extensive measurements of this multiple-wave structure in Ge [93]. The Hugoniot elastic limit (HEL) is a strong function of crystal orientation. Their data for three different crystal orientations are shown by the red circles in Fig. 18, each point being an average over 10-12 experiments. A von Mises strength model, with a shear modulus of 56.5 GPa and yield strength of 3.7 GPa, gives reasonable agreement with the average of these three points. A slight decrease in the strength after yielding was found necessary to match the state of the first plastic wave. This behavior can be reproduced by using the SGL model [15][16] with CTH parameter BST=-35.0. The calculations shown in Fig. 18 were made using this constitutive model. (The tensile strength is discussed at the end of this section.)

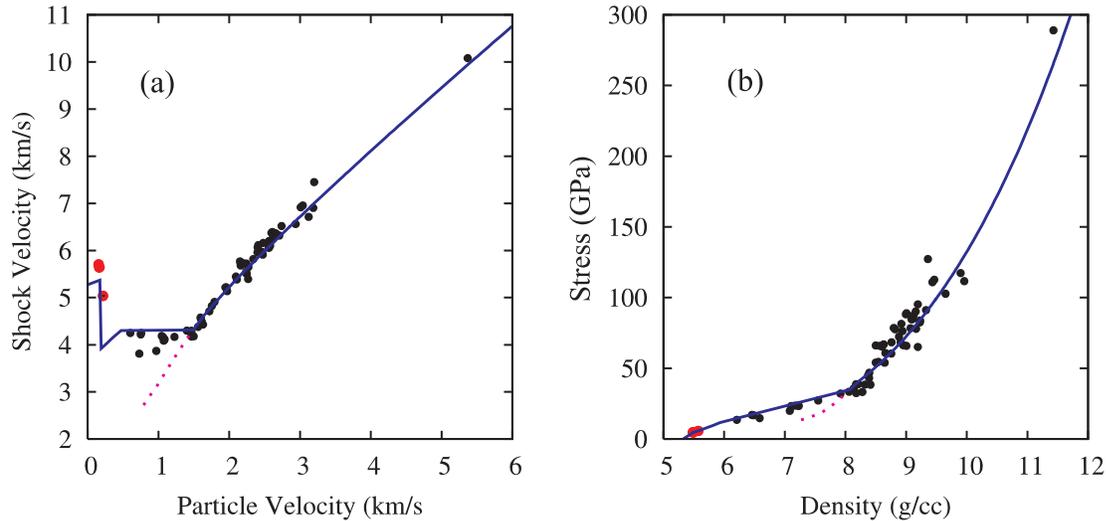

**Fig. 18. Hugoniot for germanium. Black circles are from Refs. [31][71][92]. Red circles are averages of HEL data from Ref. [93]. Solid blue curves are calculated HEL and plastic waves. Magenta curves are second plastic wave in three-wave region.**

Figure 18 shows that the calculated Hugoniot is slightly softer than the single experimental data point near 300 GPa [92]. Given the scarcity of data, it was not deemed worthwhile to adjust the model parameters to match this one point. As a test, however, the calculation was repeated using the Inferno model of Liberman [94] for the thermal electronic term in place of the simple TFD model. This substitution gave virtually perfect agreement with the data point. (Unfortunately, Liberman's model could not be used in making the final table because it must be combined with another model to span the entire range of densities and temperatures.) If nothing else, this exercise demonstates that thermal electronic excitation is an important part of the EOS model under impact conditions relevant to SPP.

Figure 19 shows plots of the pressure as a function of density on eleven isotherms, equally spaced from 250 to 1250K. The blue curves correspond to the lower temperatures, 250-650K, below the I-II-liquid triple point; they are closely spaced and exhibit the I-II solid phase transition at about 10 GPa. The red curves correspond to the higher temperatures, 750-1250K, above the triple point; they exhibit the melting transition, which is a strong function of temperature. As noted above, the liquid phase is more dense than the solid, at pressures up to the triple point, and less dense than the solid at higher pressures. The result of this behavior is that the isotherms cross at low pressures. In particular, note that the higher temperature red curves give lower pressures than the lower temperature blue curves up to 10 GPa. The pressure decreases on heating at constant density.

The behavior seen in Fig. 19 may appear unusual at first glance, but it is a natural *and unavoidable* consequence of the negative slope of the melting curve and does not violate any thermodynamic laws. The energy, however, is a monotonically increasing function of temperature, as required by thermodynamics.

There are several improvements that could be made to the new EOS. The most important one would be to replace the simplistic TFD model with a more sophisticated treatment of thermal electronic excitation and ionization. The tools for doing that are available, but their application would take some time. Additional solid phases could also be added to the model.

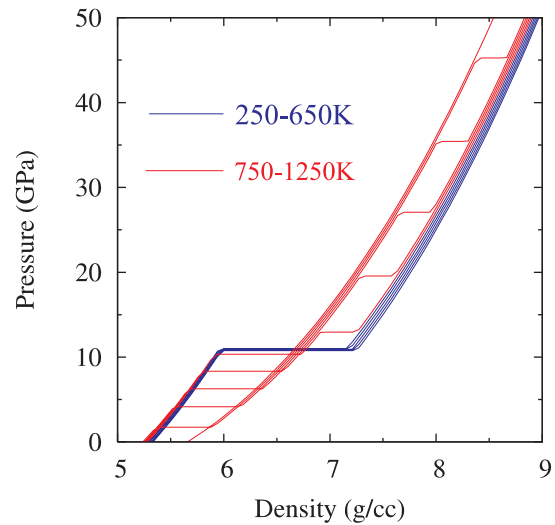

**Fig. 19. Pressure vs. density on isotherms for Ge. Blue curves correspond to temperatures from 250-650K, red curves to temperatures from 750-1250K.**

The spall strength of Ge is also needed for the numerical simulations. Claeys and Simoen [95] state that Ge is brittle at room temperature, having a tensile strength of only 40-95 MPa, but that it becomes ductile when heated to 500K. In that case, a higher value might be needed for simulations of impacts on the hot solar cells. Sensitivity studies should be made to examine the effect of varying the tensile strength. Further study of the tensile behavior may be necessary.

# 11.RECOMMENDATIONS FOR EXPERIMENTS

This report has addressed dust impact analysis primarily from a theoretical point of view, using experimental data already available in the literature. Experimental studies were not included in the statement of work, and there would not have been enough time to plan and carry them out in any case. But theoretical work, even work of the highest quality, should always be tested by comparison to experiment—whenever it is possible to do so in a meaningful way. This section will offer some recommendations for experiments to be pursued in follow-on work.

In designing experiments, it is natural to think first of studying impacts similar to those that will be experienced by components of the SPP spacecraft—accelerating 100 μm particles to velocities of 100 km/s, impacting them on mockups of the TPS and solar cells, and observing the results. Unfortunately, it is not possible to carry out such experiments using existing technologies.

It will be possible to carry out tests of this type at much lower velocities, on the order of 10 km/s. In fact, some of these tests are already being carried out. These experiments will no doubt produce some useful information. However, one should not try to extrapolate those results to higher velocities without the use of the hydrocode simulations. How then, does one test the validity of those simulations under the conditions of interest?

It may also be possible to carry out some experiments at much higher velocities, using particles accelerated by Van de Graaf generators, as described by Ratcliff, et al. [96]. However, those experiments present other problems—characterization of the particle size and impact velocity, limitations on the choice of particle type and target material, and issues relating to the charge on the particle. Even so, some experiments of this type could also be useful.

A third option is to exploit the information that has been obtained from materials recovered from space [69][97], as mentioned in Sec. 8. These data are also difficult to use as tests of numerical simulations, because information about the particle size and velocity are not available. However, further study of these data could also be useful.

I would like to offer a different way of attacking this problem, one that should be used in addition to, not in lieu of, other approaches. It is clear that hydrocode simulations will have to play the central role in understanding the effects of dust impacts. If so, the primary goal of experiments should be to help make those simulations as accurate and reliable as possible. There are two ways to accomplish that goal:

- First, by studying the material response behavior at pressure-temperature states relevant to dust impacts, generating information that can be used to construct the EOS and constitutive models.

- Second, by testing the hydrocode simulations under pressure-temperature loading conditions similar to those encountered in dust impacts, without having to reproduce the impacts.

## 11.1 Model Development

In order to obtain experimental data for use in developing material models, it is important to remember that a 100 km/s impact will produce all possible shock states in the target, i.e., the entire Hugoniot, and also most release states from those shock states. Standard quasi-static compressive and tensile strength measurements of materials provide only a tiny fraction of the information needed. It is also essential to look carefully at the areas of greatest uncertainty, not just focus on the topics of greatest interest to the individuals involved in the test program.

The present study has uncovered many gaps in the experimental data needed for developing material models needed for the simulating SPP dust impacts. I will not try to enumerate them all here, because it would be naïve and unrealistic to expect the SPP program to fill all of those gaps. I will only mention a few areas.

**Glass impact tests**. It would be very useful to have a systematic study of glass impacts, to study the crater formation and spall damage as a function of velocity under controlled conditions. The geometry should be as simple as possible, and the same material—fused silica—should be used for the impact particle and target in all tests. The goal of these tests would be to study material response, not to simulate SPP dust impacts. If these tests proved to be useful, other tests with different materials and geometries could be made later.

**Effect of glass composition on material properties**. Since the coverglass for the SPP solar cells is not going to be fused silica, it will be necessary to develop models for predicting the dependence of EOS and constitutive properties on the glass composition. The impact tests described above would be one way of studying the composition effects. If possible, it would also be useful to obtain shock-wave measurements for several compositions, over as wide a range of shock pressures as possible.[1]

**Information about DC93-500**. There are too many gaps in our knowledge of the adhesive DC93-500, discussed in Sec. 9. The elemental composition and heat of formation should be obtained from Dow Corning, so that one can assess the dif-ferences between this material and Sylgard. If these data do not already exist, they should be measured. More measurements on Sylgard itself would also be helpful.

---

1. Some data may already be available. An extensive survey of the literature should be made in conjunction with plans for new measurements.

It is also important to know more about the behavior of DC93-500 at the working temperature of the solar cells, 473K. The density of the material at this temperature is not known. The little information that has been given to us even hints that the material begins to outgas at lower temperatures, an indication of decomposition. Hence this information is not only needed for developing material models but also for insuring a design that will work properly at the elevated temperature.

## 11.2 Tests of Hydrocode Simulations

A hypervelocity dust impact is essentially an explosion. A large amount of energy is deposited into a small volume of material in a short period of time, resulting in violent ejection of some of the material and propagation of shock and release waves into the surrounding material. In order to test the ability of a hydrocode to describe such an event, it is more important to generate the right amount of energy deposition than to produce the explosion by a particle impact.

The kinetic energy of a 100 μm fused silica particle, moving at 100 km/s, is about 6 Joules. It should be possible to generate an explosion of this energy by laser deposition. Successful laser experiments would require control of the deposition time and size of the deposition region, and good diagnostics for the amount of energy deposited as well as the effects of the explosion on the target. Hydrocode calculations would be useful in designing these experiments and comparing them to dust impact events.

It should also be noted that the best experimental tests are not necessarily those that duplicate the geometry of the TPS or the solar cells. There is much to be said for using simpler geometries to facilitate the analysis and minimize opportunities for problems. If tests with simpler geometries were successful, more complicated geometries could then be used.

Finally, it should be noted that hydrodynamic scaling (Sec. 2.5) could be used in designing these experiments. If it is not possible to deposit 6 Joules of energy in a 100 μm diameter spot, one could use a larger diameter—provided that the amount of energy were increased by the cube of the diameter ratio.

Even if the explosive conditions of a dust impact cannot be duplicated exactly with lasers, it should be possible to come close enough to obtain useful information. Development of a technique for using lasers to study impact phenomena could have applications to a wide variety of problems.

# 12.SUMMARY AND CONCLUSIONS

This report discusses the EOS and constitutive properties of materials that are needed for hydrocode simulations of dust impacts on the solar probe spacecraft. New or improved material models are presented for six materials—aluminum oxide ($Al_2O_3$), a carbon-carbon composite, a very low-density carbon foam, fused silica ($SiO_2$), a silicone elastomer adhesive, and germanium. These materials will be used in the TPS and solar cells, the two spacecraft components most vulnerable to dust impacts. The report also discusses hydrocode simulations of dust impacts on the TPS and on glass targets, the importance of radiation terms in the EOS, and the possibility of reactions between carbon and $Al_2O_3$ during dust impacts.

This project has been very successful and productive. The most important material modeling issues have been identified and addressed, and a great deal has been accomplished in a remarkably short time. But it must also be said that much of this work is preliminary and that many issues need further study.

At the risk of repetition, it must be remembered that the numerical simulation of hypervelocity impacts requires "global" material models that describe all shock and release states over a phenomenal range of conditions, from very low to very high stresses and temperatures. These models cannot be constructed simply by fitting experimental data because it is not possible to make the necessary measurements over most of this range. Therefore, the credibility and reliability of the hydrocode simulations depend on the use of sophisticated theoretical models, and a high standard should be applied when evaluating them.

The problem of radiation contributions to the EOS has been solved (Sec. 3). These terms have been added to the EOS models for all materials and can be included in all the simulations. The EOS corrections appear to be quite small for impacts at the conditions anticipated. However, the present study does not address the problem of radiation emission and transport.

The new EOS for $Al_2O_3$ (Sec. 4) is a significant step forward, when compared with the models available at the beginning of this project. The new model includes a treatment of melting, dissociation (to atomic Al and O), and electronic excitation and ionization. However, the treatment of the liquid phase is very preliminary and does not reproduce some of the available data, especially the large increase in liquid density that occurs on melting. This and other features will likely require development of a chemical equilibrium model that includes formation of various Al-O compounds in the liquid phase. Models for these oxides will also be needed if it is decided to pursue further study of carbon-$Al_2O_3$ reactions (Sec. 7). Additional work on the shear and tensile properties of $Al_2O_3$ may also be needed, to determine the extent of spall damage done to this material during an impact.

A very good EOS for carbon is available for use in modeling the C-C composite and RVC foam layers of the TPS (Sec. 5). The biggest uncertainties in the models for these two materials are related to their heterogeneous microstructures. Preliminary models for this microstructure have been developed and tried in hydrocode simulations (Secs. 5 and 6). Calculations for the composite show that the microstructural effects are small, but not necessarily negligible; the crater size is somewhat larger when these effects are included, and there is some extra heating and jetting of the softer matrix material. Calculations for the RVC foam show that it offers very little resistance to penetration by ejecta from higher layers, but that there is a possibility of lateral expansion of hot gases through the open pore structure. Additional hydrocode calculations are needed to examine these phenomena, and three-dimensional simulations may be needed.

Development of a satisfactory model for the solar cell coverglass, a particularly vulnerable part of the spacecraft, has proven to be the most difficult and challenging task in the entire study (Sec. 8). A sophisticated EOS for fused $SiO_2$ was available from a previous study, but several aspects of the model—the stishovite transition, melting behavior, and dissociation—need to be improved. Even more important, there is as yet no way to develop an EOS for the real coverglass material, which will be a mixture of $SiO_2$ with six or more other metal oxides. The shear and tensile properties of glass, which play the primary role in determining the extent of spall damage around the crater, also require much more study; the model discussed in this report is very preliminary. Improvements to the EOS and constitutive properties of glass (Sec. 8.4) will require both theoretical and experimental work, a greater commitment of time and effort than any of the other follow-on studies.

A completely new EOS table has been developed for the solar cell adhesive DC93-500 (Sec. 9), using a sophisticated model that accounts for shock-induced decomposition of the material. The main weakness of this model is that it is actually an EOS for a surrogate material, Sylgard, since it was not possible to obtain detailed information for the real adhesive. There are also questions about the thermal expansion and possible decomposition of the material at the high working temperature of the solar cells. More information about the material needs to be obtained from the manufacturer. Further improvements to the EOS are not proposed at this time but could be needed later.

A completely new EOS table has also been developed for germanium (Sec. 10) using a sophisticated multiphase model that accounts for a solid-solid transition, with a 20% change in density, melting, and electronic ionization. Improvements to the model would include the use of a more sophisticated treatment of ionization equilibrium and the addition of other solid phases. Additional work on this EOS is not proposed at this time but could be needed later.

As explained in Sec. 11, the present study has been primarily theoretical. An experimental program is needed to complement these and future theoretical studies. The goals of the experiments should be two-fold—to provide data that can be used to improve the material models, and to test the ability of the hydrocode/material model package to simulate explosive events like those obtained in dust impacts.

It will not be possible to make useful measurements of dust-impact events like those that will occur on the SPP spacecraft, and a number of compromises will have to be made. I strongly urge that laser experiments be considered for testing the hydrocode capabilities. Since a hypervelocity dust impact is essentially an explosion, it should be possible to use lasers to generate conditions similar to those of an impact, i.e., by depositing energy into targets with roughly the same time scale and spatial distribution. If laser experiments are carefully designed, and carried out with suitable controls and diagnostics, they will be able to test the code simulations in a way that no other technique can do.

Equations of State," Sandia National Laboratories report SAND88-2291, 1991.

# Appendix A

# MODEL OF A C-C COMPOSITE

This appendix constructs a model for the structure of a carbon-carbon composite that is used in the CTH hydrocode simulations discussed in Secs. 5.3 and 6. The primary parameters to be determined are:

- The shape and dimensions of a fiber bundle
- The spacings between the fiber bundles in the cloth,
- The densities of the interstitial carbon matrix and the fiber bundles (after infiltration with matrix carbon).

Definitions:

| | |
|---|---|
| $d$ | diameter of a single carbon fiber |
| $W$ | width of a fiber bundle (dimension in plane of cloth) |
| $t$ | thickness of a bundle (dimension perpendicular to plane) |
| $q$ | spacing between bundles (in plane) |
| $\rho_C$ | average density of composite |
| $\rho_F$ | density of a single carbon fiber |
| $\rho_M$ | density of interstitial carbon matrix |
| $\rho_B$ | density of fiber bundle after infiltration with carbon |

Assumptions: Some of the parameters needed for the model are not known, and reasonable estimates must be made. The following typical values are used.

- Fiber diameters are typically 5-10 µm. $d = 7$ µm is used here.
- Fiber bundles typically contain 1000, 3000, or 5000 fibers. 1000 fibers per bundle is the value used here.
- Bundles are taken to have an elliptical cross-section.
- The bundle thickness $t$ can be estimated from the number of cloth layers in a given thickness of the composite. A thickness of 69 µm is used here.
- Fibers in a bundle are assumed to be randomly packed, giving a void area of 17% in a bundle cross-section.
- The cloth fabric is assumed to be a 4-harness satin weave.
- The spacing between bundles is assumed to be twice the bundle thickness, as discussed below.
- The composite density, $\rho_C$, is assumed to be 1.57 g/cm$^3$.
- The density of an individual fiber, $\rho_F$, is assumed to be 2.1 g/cm$^3$.

These assumptions are consistent with the information I have been able to obtain and are sufficiently representative for the purposes of the present study.

First consider the dimensions of a fiber bundle. Assuming 1000 fibers per bundle, randomly-packed, the dimensions of a bundle can be computed from the fiber diameter as follows:

$$1000\pi d^2/4 = 0.83\pi Wt/4, \; W = 1205 d^2/t = 855 \text{ μm},$$ (A.1)

using the values of $d$ and $t$ given above.

The next step is to calculate the volume fraction occupied by the fiber bundles in the cloth fabric. The "unit cell" for a 4-harness satin weave cloth, as described in Ref. [58], consists of two thickness of bundles, oriented perpendicular to one another, and containing four bundles in each direction. The volume of a unit cell is thus seen to be

$$V = 2t(4W + 4q)^2 = 32W^2 t(1 + q/W)^2 .^1$$ (A.2)

If the two bundle layers lay one on top of another, the length of a bundle within a unit cell would be $4W + 4q$. However, each bundle crosses three bundles in the other layer and then wraps around the fourth, as shown in Fig. 4 of Ref. [58]. Adding a small correction to the length, one finds the volume occupied by the eight bundles to be

$$V_B = 8(4W + 4q)(1 + \delta)\pi Wt/4 .$$ (A.3)

The volume fraction occupied by the bundles is

$$v_B = V_B/V = (\pi/4)((1 + \delta)/(1 + q/W)) .$$ (A.4)

The spacing $q$ must be greater than the thickness $t$, in order to allow the bundles in one layer to wrap around those in the other layer. The larger the value of $q$, the more flexible the cloth will be. In the absence of actual measurements, it is reasonable to try $q \sim 2t = 138$ μm.

I have not attempted an exact calculation of the length correction $\delta$, but a rough estimate gives $\delta \sim 0.01$ for the values of $q$, $t$, and $W$ used here. This correction is negligible in comparison with the uncertainty in $q$ itself. (In fact, this correction is not needed for the simple model discussed in Sec. 5.3, in which the fibers do not wrap around those in different layers.)

---

1. This result differs slightly from that given in [58], where it was assumed that a unit cell included only three spacings between bundles. It is necessary to include the same number of spacings as bundles so that a collection of unit cells fills all space.

The next step is to compute the volume fraction occupied by the individual fibers in the cloth, by multiplying $v_B$ by the packing fraction within a bundle. Using the values of $q$, $t$, and $W$ obtained above,

$$v_F = 0.83 v_B = 0.83 (\pi/4)/(1 + 2t/W) = 0.56 \qquad (A.5)$$

According to this estimate, 56% of the composite volume consists of carbon fibers, having a density close to that of graphite, and 44% consists of interstitial carbon. These values are in rough agreement with those quoted by manufacturers.

The density of the interstitial carbon matrix can be determined from the fiber volume fraction , the average composite density, and the fiber density, using

$$1/\rho_C = v_F/\rho_F + (1 - v_F)/\rho_M. \qquad (A.6)$$

Taking $v_F = 0.56$, $\rho_A = 1.57$, and $\rho_F = 2.1$, one obtains $\rho_M = 1.19$. Hence the interstitial carbon is itself a porous material, the pores being much smaller than the spaces between the fiber bundles.

Finally, the average density of a fiber bundle, $\rho_B$, *after* infiltration with the carbon matrix, is then seen to be

$$1/\rho_B = 0.83/\rho_F + 0.17/\rho_M \ , \ \rho_B = 1.86. \qquad (A.7)$$

Hence the fiber bundles are also porous, though not as porous as the matrix.

The most important results are summarized below:

- The fiber bundles have an elliptical cross-section, 69 μm in thickness and 855 μm in width.
- The spacings between the fiber bundles in the cloth is 138 μm.
- The density of the interstitial carbon matrix is 1.18 g/cm$^3$.
- The density of the fiber bundles, after infiltration with matrix carbon, is 1.86 g/cm$^3$.

# Appendix B

# MODEL OF RVC FOAM

As noted in Sec. 5.4, an RVC foam is characterized by open cells, or bubbles, formed by a network of filaments, or struts. This appendix first discusses a three-dimensional model of this structure that could be used in hydrocode simulations. It then presents a highly simplified two-dimensional model that is actually used in the hydrocode calculations discussed in Sec. 6.

**Three-Dimensional Model**. An RVC foam can be conceived as a random packing of bubbles, or cells, the struts forming the intersections between them. The principal features of this structure can be obtained by placing the bubbles on a regular body-centered cubic (bcc) lattice. In that case, each cell will be identified with the rhombic dodecahedron that is formed from the planes bisecting first- and second-nearest neighbors. Each polyhedron has fourteen faces; eight are octagonal (nearest neighbors), and six are square (next-nearest neighbors). Each polyhedron has 36 equal edges, which represent the struts in the foam.

The following results can be derived from the geometry of the rhombic dodecahedron, described in Refs. [98]. The length of an edge $a$ is related to the distance between cells $D$ by

$$a = D/\sqrt{6} = 0.40825D. \tag{B.1}$$

The volume of the polyhedron is given by

$$V = 8\sqrt{2}a^3 = (4/3\sqrt{3})D^3. \tag{B.2}$$

Now represent the struts by cylinders of diameter $s$, lying along the 36 edges of the polyhedron. Let $V_s$ be the volume occupied by the struts. It can be shown, using the dihedral angles, that one-third of the volume of each strut lies within a cell.[1] For purposes of computing the volume, the strut length is approximately $a - s/2$, after correcting for overlap. Hence the strut volume is

$$V_s = 36(1/3)(\pi s^2/2)(a - s/2). \tag{B.3}$$

The strut volume fraction is

---

1. The dihedral angle is the angle formed by two adjacent planes in the polyhedron. In a rhombic dodecahedron, 24 edges correspond to an angle of 125.27 and 12 edges correspond to an angle of 109.47, giving an average of 120.

$$v_s = V_s/V = (3\pi/8\sqrt{2})x^2(1-x/2) = 0.83304x^2(1-x/2), \qquad \text{(B.4)}$$

where $x = s/a$.

Now consider a 100 ppi 3% RVC. Taking $v_s = 0.03$, Eq. (B.4) gives a strut diameter $s = 0.2a$. As noted in Sec. 5.4, the bubble spacing corresponding to 100 ppi is 350-500 µm. Taking $D = 425\,\mu m$, one obtains $a = 174\,\mu m$, $s = 35\,\mu m$. The length of a strut, as it would appear in SEM, is $l = a - s = 139\,\mu m$. These results are in rough agreement with the values $l = 130$, $s = 50$, obtained using scanning electron microscopy (SEM) on 100 ppi 3% RVC [62].

**Two-Dimensional Model**. Use of the above model would require a very expensive three-dimensional hydrocode calculation. Therefore, a much simpler model was also developed for use in the preliminary and exploratory computations discussed in Secs 5.4 and 6.

In order to set up the calculation in two-dimensional, cylindrical geometry, the polyhedra cells were replaced by toruses having a hexgonal cross-section. The basic geometry is depicted in Fig. 20.

Each cell is represented by a hexagon with edge length $a$. The distance between hexagons is given by

$$D = \sqrt{3}a. \qquad \text{(B.5)}$$

Letting $s$ be the width of the struts (shown in blue), the void region is described by a smaller hexagon with edge length

$$a_0 = a - s/\sqrt{3}. \qquad \text{(B.6)}$$

The strut volume fraction can be computed from the areas of the inner and outer hexagons. The result is

$$v_s = 1 - (a_0/a)^2. \qquad \text{(B.7)}$$

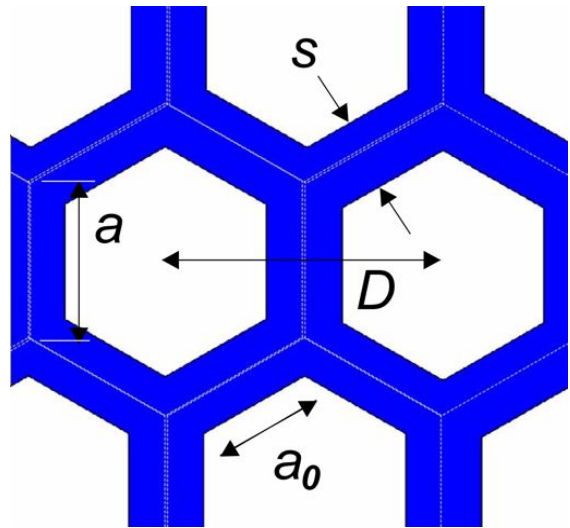

**Fig. 20. Cross-sectional structure of RVC foam in 2-D hydrocode model.**

Now consider a spacing $D = 425\,\mu m$ (for 100 ppi) and strut diameter $s = 50\,\mu m$ (from SEM data). The corresponding strut length is $a_0 = 216\,\mu m$. This result is 66% larger than the value found by SEM, but close enough for use in exploratory calculations.

A more serious problem with the model is that the strut volume fraction, $v_s = 0.22$, is much too large. The primary reason for this discrepancy is that the toruses are completely closed on the sides; the model does not reproduce the open pore structure that extends in all directions in the real material.

Hence it is necessary to reduce the density of the carbon in the struts from 1.65 to 0.23 g/cm$^3$. In other words, the 2-D model only accounts for the pores in one direction, that parallel to the sides of the torus. Pore compaction in the other two directions must be accounted for using the p-alpha model. See Sec. 6 for further discussion.

# Appendix C

# EOSPRO INPUT FILES FOR GERMANIUM

The following EOSPro input files were used to generate the three-phase EOS for germanium discussed in Sec. 10. These files will require some editing to be used with the older Panda code.

```
!**********************************************************************
!
!  01/14/09 - EOS for Ge-I (cubic diamond), EOS number 101.
!  Cold curve: Birch-Murnaghan formula, R0=5.3598, B0=76.5, BP=4.0,
!     parameters chosen to fit RTP density and sound speed, and to
!     give reasonable extrapolation to TFD curve at high pressures.
!  Lattice vibrations: Einstein model.
!  --Gruneisen parameter from thermodynamic data.
!  --Einstein temperature chosen to match entropy data for solid.
!  Electronic term: None. Material is an insulator.
!  Energy shift: esft is chosen to give zero enthalpy at RTP.
!
!**********************************************************************
reset
! Setup
sym  z=32  aw=72.610  matid=101  bfile='b101'  afile='a101'
sym  ecoh=5.12  rref=5.327  gref=0.8085  dref=240
sym  rzro=5.3599  bzro=76.5  dbdp=4.0 rtfd=10.0
sym  esft=-0.10814  r0=5.3270  r1=5.3622  r2=18  r3=110
!
mod sol crv=1 nuc=1 esft=esft
Ge[1]
ecoh rref 298 gref dref
6 rtfd
rzro bzro dbdp
-3 2 - - -
isob sol 0 298 0 1 rref
! Make EOS table - note limits on mesh
mesh sol
0.95*r0 r0 5 1
r1 r2 17 3 298      ! 17 points from .5 to 1000 GPa
r2 r3 7 3 298       ! 7 points from 1.e3 to 1.e5 GPa
500 0 1 1
0 0 0 0
5 200 3 1
298 300 2 1
400 2000 17 1
2000 1.e4 15 2
0 0 0 0
slib sol
201
z aw 0 298 rref
301
no
0
matid bfile afile
end
!**********************************************************************
!
!  01/14/09 - EOS for Ge-II (beta-Sn), EOS number 102.
```

```
!  Cold curve: Birch-Murnaghan expression, R0=6.57, B0=77., BP=4.5,
!    parameters chosen to fit RT isotherm up to ~75 GPa, where
!    transformation to higher-density phases occur, and to agree with
!    Hugoniot data when cold curve is used in liquid model.
!  Lattice vibrations: Debye model with cutoff.
!  --Debye temperature DREF=320 and Gruneisen parameter GREF=2.0
!    chosen to give nearly vertical phase boundary with Ge-I.
!  Electronic term: TFD.
!  Energy shift: esft is chosen to match phase transition pressure
!    of 10.8 GPa at RT, assuming zero enthalpy for Ge-I at RTP. The
!    value given here is ~0.34 MJ/kg higher than for Ge-I, which is
!    close to the value 0.33, given by Chang & Cohen.
!
!*********************************************************************
reset
! Setup
sym  z=32  aw=72.610  matid=102  bfile='b102'  afile='a102'
sym  ecoh=5.12  rref=6.447  gref=2.0  dref=320
sym  rzro=6.57  bzro=77  dbdp=4.5  rtfd=15.0
sym  esft=-.11091+0.34  r0=6.4467  r1=6.4916  r2=19  r3=110
!
mod sol  crv=1 nuc=1 tel=1 esft=esft
Ge[1]
ecoh rref 298 gref dref
6 rtfd
rzro bzro dbdp
-1 3 - - -
tfd
isob sol 0 298 0 1 rref
! Make EOS table
mesh sol
0.95*r0 r0 5 1
r1 r2 17 3 298      ! 17 points from .5 to 1000 GPa
r2 r3 7 3 298       ! 7 points from 1.e3 to 1.e5 GPa
500 0 1 2
0 0 0 0
5 200 3 1
298 300 2 1
400 2000 17 1
2000 1.e4 15 2
0 0 0 0
slib sol
201
z aw 0 298 rref
301
no
0
matid bfile afile
end
!*********************************************************************
!
!  01/15/09 - EOS for liquid Ge, EOS number 110.
!  Cold curve: Birch-Murnaghan expression, R0=6.57, B0=77., BP=4.5,
!    same parameters as used for beta-Sn phase, Ge-II.
!  Electronic term: TFD.
!  Cohesive energy: Using heat of formation of Ge gas. This value,
!    together with parameter EFAC, determines vapor pressure.
!    Note: Set parameter DREF to small value so that EBZPE=ECOH.
!  Energy shift: esft same value as for Ge-II, which assumes zero
!    enthalpy of Ge-I at RTP.
!  CRIS Model: Parameter EFAC adjusted to match free energy of Ge-I
!    at melting point. Parameters WX1 & WX2 same as for carbon and
!    silicon. Statistical weight of 4 needed to match entropy at
!    melting point.
```

```
!  Vapor-Liquid coexistence region: Experimental b. p. is 3106K.
!    Model predicts pv=1.0e-4, rhov=2.2e-4, rhol=5.15 at 2868K.
!    Model predicts critical point at 8268K, 0.76 GPa, 1.66 g/cc.
!
!*********************************************************************
reset
! Setup
sym  z=32  aw=72.610  matid=110  bfile='b110'  afile='a110'
sym  ecoh=5.12  rref=6.447  gref=2.0  dref=0.01
sym  rzro=6.43  bzro=69  dbdp=4.5  rtfd=15.0
sym  rlj=0.90*rzro flj=0.75  efac=.0334*ecoh  wx1=1.25  wx2=1.25
sym  esft=-.11069+0.34  r0=5.327  r1=5.3622  r2=19  r3=110
!
mod liq  crs=1 tel=1 esft=esft wlq=4
Ge[1]
ecoh rref 298 gref dref
6 rtfd rlj flj
rzro bzro dbdp
0 efac 0 0 1000 2 - - - - wx1 wx2
tfd
isob liq 1.e-4 298 0 1 1.3e-3
isob liq 0 1211.4 0 1 5.6
vap liq
3106 0 1 6.e-4 5.3

! Make EOS table
mesh liq
0 0 1 1
1.e-10 1.e-6 5 1
1.e-6 .05 10 2
.05 .50 5 2
0.50 r0 25 1         ! 25 points for decade below ref. point
r1 r2 50 2 298       ! 50 points from .5 to 1000 GPa
r2 r3 7 3 298        ! 7 points from 1.e3 to 1.e5 GPa
200. 0 1 1
0 0 0 0
5 200 3 1
298 300 2 1
400 3000 27 1
2866 2870 5 1
3000 7400 20 2
7400 8200 9 1
8300 9300 3 1
8200 1.e5 15 2
1.e5 1.e8 16 2
0 0 0 0
save mesh mshliq
slib liq
201
z aw 0 298 rref
301
no
0
matid bfile afile
end
!*********************************************************************
!
!  01/16/09 - Multiphase EOS for Ge, EOS number 3820.
!  Two solid phases and a fluid phases are included.
!    EOS #101 - Ge-I (diamond)
!    EOS #102 - Ge-II (beta-Sn)
!    EOS #110 - fluid
!  Maxwell Constructions: TSPALL=3106K, RSPALL=?, TCRIT=?K.
!
```

```
!*******************************************************************
reset
!
sym rref=5.327 z=32 aw=72.61 matid=3820 bfile='b3820' afile='a3820'
!
set bas
Ge
- rref 298
mod trn  rref=rref
 matid=101 file=b101 name=Ge1
 matid=102 file=b102 name=Ge2 !eshift=-0.05
 matid=110 file=b110 name=Liq liq=1

! Make EOS table
read mesh mshliq
slib trn
201
z aw 0 298 rref
301
yes
2868 5.15
2868.5 8268 2.2e-4 5.15
1.1444 2.2060
1.3080 2.0216
-
-
matid bfile afile
end
```